\documentclass{aa}
\usepackage{hyperref}
\usepackage[varg]{txfonts} 
\usepackage{natbib}
\usepackage{graphicx}
\usepackage{amsmath}
\usepackage{ulem}
\usepackage{tabularx}
\usepackage{comment}
\usepackage[dvipsnames]{xcolor}
\begin{document}

\newcolumntype{L}{>{\raggedright\arraybackslash}X}
\newcommand{\sepp}{\texttt{SourceXtractor++ }}
\newcommand{\se}{\texttt{SExtractor }}

\title{Structural evolution of quiescent galaxies at intermediate redshifts at UV and red rest-frame wavelengths} 

\author{Michele Pizzardo \inst{\ref{1},\ref{2}}\thanks{\email{mpizzardo@asiaa.sinica.edu.tw}}
\and Ivana Damjanov \inst{\ref{1}}
\and Jubee Sohn \inst{\ref{3}}
\and Margaret J. Geller \inst{\ref{4}}
}

\authorrunning{M. Pizzardo et al.}

\institute{\label{1}Department of Astronomy and Physics, Saint Mary's University, 923 Robie Street, Halifax, NS-B3H3C3, Canada
\and \label{2}Academia Sinica Institute of Astronomy and Astrophysics (ASIAA), No. 1, Section 4, Roosevelt Road, Taipei 106216, Taiwan
\and \label{3}Astronomy Program, Department of Physics and Astronomy, Seoul National University, Gwanakgu, Seoul 151-742, Republic of Korea
\and \label{4}Smithsonian Astrophysical Observatory, 60 Garden Street, Cambridge, MA-02138, USA
} 

\date{Received date / Accepted date}

\abstract 
{
We derived structural parameters for a mass-limited sample ($M_\star>10^{10}\, M_\odot$) of $~27,000$ quiescent galaxies with $0.2 < z < 0.6$ using $grizy$ photometry from Subaru/Hyper Suprime-Cam and dense spectroscopy from the HectoMAP survey. Based on S\'ersic profile fits in all five bands, we modeled the wavelength dependence of the circularized half-light radius ($R_{e,c}$, a proxy for size) and S\'ersic index ($n$, a proxy for central concentration). We estimated the structural parameters in two rest frames: UV (3500\, \AA, tracing young and metal-poor stellar populations) and red (7000\, \AA, tracing the stellar mass distribution). Combined with the stellar mass, redshift, and D$_n4000$, the estimates of $R_{e,c}$ and $n$ enable us to explore the evolution in the correlations between structural properties and stellar mass for quiescent galaxies with different stellar population ages. At intermediate redshift, quiescent galaxies at all stellar masses exhibit a systematic decline in size and a rise in central concentration with the rest-frame wavelength.
Over the full redshift range of the sample,  variations in the Sérsic index are larger for galaxies that recently joined the quiescent population (newcomers, with $1.5 <$ D$_n4000 < 1.6$) than for descendants of galaxies that are already quiescent at the survey limit, $z \sim 0.6$ (aging, or resident, population). Variations in size with the rest-frame wavelength are greater for newcomers than for the resident population, with $ z < 0.3$. The combined evidence supports inside-out quenching as the dominant mechanism halting star formation during this epoch. The normalization of the size-stellar mass relation (the typical size of a $M_\star\sim10^{11}\, M_\odot$ quiescent galaxy) increases by $\sim30\%$ between $z \sim 0.6$ and $z \sim 0.2$ in the red rest frame and remains constant in the rest-frame UV. Size growth is age-independent, and newcomers are found to be $\sim20\%$ larger than the aging population across redshifts and rest frames. Based on the S\'ersic index-stellar mass relation in the UV, we find that $M_\star \sim 10^{11}\, M_\odot$ quiescent galaxies maintain a constant central concentration ($n\sim4$ for the aging population and $n\sim2$ for newcomers). In the red rest frame, both subpopulations exhibit de Vaucouleurs profiles after accounting for selection effects. The larger sizes and disk-like UV profiles of newcomers link them to their direct progenitors in the star-forming population.  For the aging population, the steady red rest-frame growth and elevated central concentration in both red and UV rest frames suggest minor mergers with progressively redder satellites at lower redshifts as the primary driver of galaxy evolution in the quiescent phase. Large samples of low-surface-brightness galaxies available in forthcoming sensitive large-area imaging surveys are ideal testbeds for this prediction. 
}


\maketitle

\section{Introduction}\label{sec:introduction} 

The observed evolution in galaxy morphological properties with redshift is a fundamental test of theories that describe galaxy mass assembly. For galaxies that have just halted their global star formation processes,  the sizes and central concentrations derived from light profiles at different rest-frame wavelengths can constrain physical mechanisms  driving their evolution. For galaxies with old stellar populations, estimates of the structural parameters at multiple fixed rest-frame wavelengths can constrain the channels for growth in the quiescent phase. 

Size growth at fixed stellar mass is the most widely studied aspect of the structural evolution of quiescent galaxies. The evaluation of the size growth of the quiescent population over a broad redshift range is based on the size-stellar mass relation, a projection of the stellar mass fundamental plane \citep[e.g.,][]{Hyde09,Bezanson13,Zahid16}. For quiescent galaxies with stellar masses $\gtrsim 2\times10^{10}$M$_\odot$, this relation takes the form of $R_e = R_0 (M_\star/M_0)^\alpha$, where $R_e$ is the radius enclosing half of the total light profile of a galaxy (i.e., half-light radius), $M_\star$ is the galaxy stellar mass, and $R_0$ is the characteristic half-light radius (a proxy for size) of galaxies with $M_\star = M_0$. Based primarily on galaxy light profiles in a single photometric band, previous investigations suggest that the slope, $\alpha$, varies very little with redshift; in contrast, the normalization, $R_0$, increases substantially. From a redshift of $\sim 2$ to the present, systems with masses $\gtrsim 2\times10^{10}$M$_\odot$ grow by a factor of 2.5-4 \citep[e.g.,][]{Daddi05,Trujillo07,vandokkum10,Damjanov11,Newman12,Huertasc13,vdWel14,Huertasc15,Faisst17,Damjanov19,Mowla19,Kawinwanichakij21,Barone22,Hamadouche22,Damjanov23,George24,EQ1morpho25}. Recent results from the James Webb Space Telescope (JWST) extend the constraints on this evolution to higher redshift \citep{Cutler24,Ito24,Martorano24,Hamadouche25,Kawinwanichakij2025}.

The evolution of the S\'ersic index, a proxy for the central concentration of galaxy light,  probes the connection between galaxy mass assembly and the distribution of its stellar content with cosmic time. Single photometric band investigations at redshift $\lesssim 2.5$ \citep{Huertasc13,Patel13,Lang14,Whitaker15,Martorano25} demonstrate that the median Sérsic index of star-forming and quiescent galaxies evolves slowly or not at all with redshift. Very-high-mass quiescent galaxies $(M_\star > 10^{11}$~M$_\odot)$ display the largest redshift dependence, with an increase from $n\sim 3.5$ to $n\sim 4.5$ as redshift decreases from $z\sim 1.5$ to $z\sim 0.5$.

Values of the morphological parameters depend on the rest-frame wavelength range corresponding to the photometric band at the redshift of the observations.  A single observed band probes different rest-frame regimes (dominated by the light from different stellar populations) at different galaxy redshifts. Addressing the evolution of structural scaling relations at a fixed rest-frame as opposed to observed wavelength is critical for tracing the changes in the extent and distribution of a single stellar population component. 

Investigators tackle the wavelength dependence of morphological parameters in different ways. Often, analyses adopt a single rest-frame wavelength to explore the bulk of the galaxy stellar mass content, typically $\sim 5000$~\AA\ \citep[e.g.,][]{vdWel14,Mowla19,Kawinwanichakij21}. Galaxy light profiles at this wavelength are dominated by the light from low-mass stars that trace the total stellar mass of a galaxy \citep[e.g.,][]{Bruzual03}.
Alternatively, multiwavelength studies use mass-weighted structural parameters based on measurements from galaxy light profiles in a fixed band multiplied by the profile of the stellar-mass-to-light ratio ($M/L$) in that band from (spatially resolved) spectral energy distribution fittings \citep[e.g.,][]{Suess19a,Suess19b,Clausen25}. Some multiband fitting algorithms provide automated extraction of morphological parameters as functions of wavelength \citep{Haubler2013,Nedkova21}.

The change in the size and shape of galaxies with rest-frame wavelengths for galaxy samples with precisely determined redshifts can trace differences in the distribution of their stellar populations as a function of different properties (age, metallicity). Red,  infrared light traces the mass content of galaxies; in contrast, ultraviolet light (UV, $\sim (2000-4000)$~\AA) probes the distribution of recently formed and/or metal-poor stars  \citep[e.g.,][]{Bruzual03}. Wavelength-related variations in the structural properties (size, central concentration) of (globally) young quiescent galaxies directly reflect the dominant physical mechanisms that halt star formation in these systems \citep[e.g., inside-out vs. inside-in quenching,][]{Avila18,Lin19,Papaderos22}. Changes in structural parameters with wavelength for galaxies dominated by the oldest stellar populations track galaxy evolution during the quiescent phase \citep[e.g., the growth in size via minor mergers,][]{Naab09,Nipoti09,Hilz2012,Oser2012,Nipoti12}.

The rest-frame wavelength dependence of galaxy morphological parameters has been studied in detail with large spectroscopic samples of quiescent systems at $z\lesssim 0.2$ \citep[e.g., GAMA,][]{Driver09,Kelvin12,Kennedy16}. At higher redshifts ($\sim 0.5-2$) the trends in the structural properties with rest-frame wavelength are available only for small spectro-photometric samples from HST \citep[e.g.,][]{Guo11,vdWel14,Suess20} and JWST \citep[e.g.,][]{Clausen25} or for larger purely photometric samples \citep[e.g., from Subaru/HSC or JWST,][]{Cutler24,George24,Martorano24,Martorano25}. A study of the first statistically large dataset from the \textit{Euclid} satellite includes measurements of morphological parameters in the visible and near-infrared filters for a photometric sample of $\sim 100,000$ quiescent galaxies at $z\lesssim1$ \citep[over $<0.5\%$ of the planned total survey area,][]{EQ1morpho25}.

These studies show that for massive quiescent galaxies ($M_\star\gtrsim10^{10}\, M_\odot$) covering a broad redshift range, the galaxy size decreases and the S\'ersic index increases with increasing wavelength. The redder light profiles of quiescent galaxies reflect the older stellar population and the bulk of the galaxy stellar mass. They cover smaller galactocentric radial ranges and are more centrally concentrated than their bluer counterparts dominated by the light from young and/or metal-poor stellar populations.

The investigation of structural parameters with a large, mass-complete sample of quiescent galaxies including both high-quality multiband imaging and spectroscopy in the redshift range $0.2<z<0.6$ is a missing link between studies with statistically large samples at $z\sim0$ and results based on much smaller sets of spectroscopically confirmed quiescent systems at $z>0.5$. Here, we provide the first measurements covering this redshift gap. We extend studies of wavelength-dependent structural parameters with statistical samples of quiescent galaxies up to $z\sim0.6$.

We selected the quiescent sample from the HectoMAP spectroscopic survey \citep{Geller15Schw,Sohn21HDR1,Sohn23}, a dense red-selected redshift survey covering 55 square degrees of the northern sky. HectoMAP includes a complete mass-limited sample of $\sim 40,000$ quiescent systems at $0.2 <z <0.6$. In addition to precise redshift measurements, medium-resolution spectroscopy in the observed $3500-9000$~\AA\ interval enables measurements of the spectral index $D_n4000$, which is a ratio of the flux in the continuum around the (rest-frame) $4000$~\AA\ break and a proxy for average galaxy stellar population age. The HectoMAP field is included in the Subaru Hyper Suprime-Cam (HSC) Subaru Strategic Program (SSP) multiband ($grizy$) imaging \citep{Aihara22}. Dense spectroscopy combined with multiwavelength high-resolution imaging makes the HectoMAP galaxy sample a unique tracer of the evolution in galaxy morphological parameters over the redshift range of $0.2<z<0.6$ as a function of rest-frame wavelength, stellar mass, and average stellar population age. 

We summarize the HectoMAP survey in Sect. \ref{subsec:hectomap} and review  the complete sample of quiescent galaxies in Sect. \ref{subsec:masscomplete}. We describe the HSC-SSP imaging in Sect. \ref{subsec:hscintro}, the Sérsic fitting procedure in Sect. \ref{subsec:sepp}, and the results of multiband fitting to the HectoMAP galaxies in Sect. \ref{subsec:ress}. We describe the resulting wavelength dependence of size and Sérsic index in Sect. \ref{subsec:grad}, and we investigate the dependence of these (anti-)correlations on $D_n4000$ in Sect. \ref{subsec:grad_dn4000}. In Sects. \ref{subsec:error} and \ref{subsec:scalingrel} we investigate how the size–mass and Sérsic index–mass relations evolve in two rest-frame regimes, mapping structural changes in different stellar populations of quiescent systems. We present our conclusions in Sect. \ref{sec:conclusion}.

\section{Spectroscopy from MMT/Hectospec}\label{sec:data}

We explored the wavelength dependence of the structural parameters of quiescent galaxies at redshifts 0.2 to 0.6 based on single-Sérsic profile fitting.
We combined medium-resolution MMT/Hectospec spectroscopy and high-resolution multiband Subaru/HSC imaging to select and characterize the sample.

Section \ref{subsec:hectomap} briefly outlines HectoMAP, the spectroscopic galaxy survey that provides the parent sample of quiescent galaxies. Section \ref{subsec:masscomplete}  describes the mass-complete sample of HectoMAP galaxies we analyze.
Table \ref{table:hectomap} lists the main characteristics of the HectoMAP parent sample along with the selection that underlies the galaxy sample we use. 

\subsection{The HectoMAP galaxy redshift survey}\label{subsec:hectomap}

HectoMAP is a large dense redshift survey of quiescent galaxies covering redshifts $0.2 < z < 0.6$ carried out with the Hectospec 300-fiber wide-field instrument on the 6.5-meter MMT \citep{Fabricant05}.
HectoMAP includes spectra for 95,403 galaxies with $r<23$ covering 54.64 deg$^{2}$ over a narrow strip of the sky at $200 <$ R.A. (deg) $< 250$ and $42.5 <$ Decl. (deg) $< 44.0$ (first and second rows in Table \ref{table:hectomap}). The median redshift of HectoMAP is $z=0.345$ \citep{Geller11Russel,Geller15Schw,Sohn21HDR1,Sohn23}.  

\begin{table}[htbp]
\begin{center}
\caption{\label{table:hectomap} Selection of the HectoMAP Subsample.}
\begin{tabular}{lc}
\hline
\hline
SDSS $r$-band magnitude limit                 & 23 \\
$N_{\rm spec,tot}$                   & 95,403 \\
Color selection (cs)                 & $g-r >1,$ for $r<20.5$\\
                                     & $g-r >1$ and $r-i>0.5$, \\
                                     & for $20.5 < r<21.3$ \\
$N_{\rm spec,cs}$                    & 74,365 \\
$N_{\rm spec,cs,M_\star}$            & 72,110 \\
$N_{\rm spec,cs,M_\star,D_n4000}$                      & 69,563$^{\bf a}$ \\
$N_{\rm spec,cs,M_\star,D_n4000>1.5}$                  & 51,274 \\
\hline
$z$ selection for mass-limited sample                  & $(0.2,0.6)^{\bf b}$ \\
$N_{\rm mass-limited\,sample}$                         & 39,468 \\
$grizy$ fits                                          & 28,128$^{\bf c}$ \\
with $R_e<0.5\times$seeing                             & 26,912 \\
\hline
\end{tabular}
\end{center}
{\bf Notes.}

$^{\bf a}$ Criteria for reliable $D_n4000$ measurements are in Sect. \ref{subsec:hectomap}.

$^{\bf b}$ We remove galaxies with $0.4<z<0.42$ because of night sky line contamination \citep{Damjanov24}.

$^{\bf c}$ HSC-SSP coverage of the HectoMAP field is $\sim 80\%$ for the $i$ and $y$ bands, and $100\%$ for the $g$, $r$, and $z$ bands (Sect. \ref{subsec:hscintro}).

\end{table}

The photometric calibration of HectoMAP survey is based on the Sloan Digital Sky Survey (SDSS) DR16 \citep{Ahumada20}. HectoMAP incudes two samples: a bright survey with $r_{petro, 0} < 20.5$ and a red color selection $(g-r)_{model, 0} > 1$, and a fainter survey with $20.5 < r_{petro, 0} \leq 21.3$ and red selection $(g-r)_{model, 0} > 1$ and $(r-i)_{model, 0} > 0.5$. The entire survey contains 74,365 galaxies (third and fourth rows in Table \ref{table:hectomap}). Here $r_{petro, 0}$ refers to the SDSS Petrosian magnitude corrected for Galactic extinction. HectoMAP color selection is based on the extinction-corrected SDSS model magnitudes.  For the faint portion, the $(r-i)_{model, 0}$ cut removes low redshift objects. There is an additional surface brightness limit  $r_{fiber, 0} < 22.0$ throughout the HectoMAP survey because red objects below this limit yield very low signal-to-noise ratio MMT spectra. This limiting $r_{fiber,0}$ magnitude corresponds to the extinction-corrected flux within the Hectospec fiber aperture (with $0\farcs75$ radius). 

\citet{Sohn21HDR1, Sohn23} described two HectoMAP data releases that include the entire redshift survey. These releases include the redshifts, spectral index $D_n4000$ measurements, and stellar masses along with the match to the SDSS photometry. We summarize these data below. 

HectoMAP includes $\sim2000$ redshifts per square degree. The full survey is $81\%$ complete at $r = 20.5$ and 72\% complete at $r = 21.3$. For red objects within the color selection limits, the completeness of HectoMAP is remarkably uniform across the area of the survey.

In \citet{Sohn23}, we computed the stellar masses for 99.8\% of HectoMAP galaxies based on SDSS DR16 $ugriz$ photometry. We used the Le Phare software package \citep{Arnouts1999,Ilbert06} that incorporates stellar population synthesis models of \citet{Bruzual03}, a \citet{Chabrier03} initial mass function, and a \citet{Calzetti00} extinction law. Following the approach from our earlier studies of galaxies at $z<1$ \citep[e.g.,][]{Geller14}, we explored two metallicities (0.4 and 1 solar) and a set of exponentially decreasing star formation rates. We derived the mass-to-light ratio from the best model and used this ratio to convert the observed luminosity into stellar mass.

\begin{figure*}
    \centering
    \includegraphics[width=\textwidth]{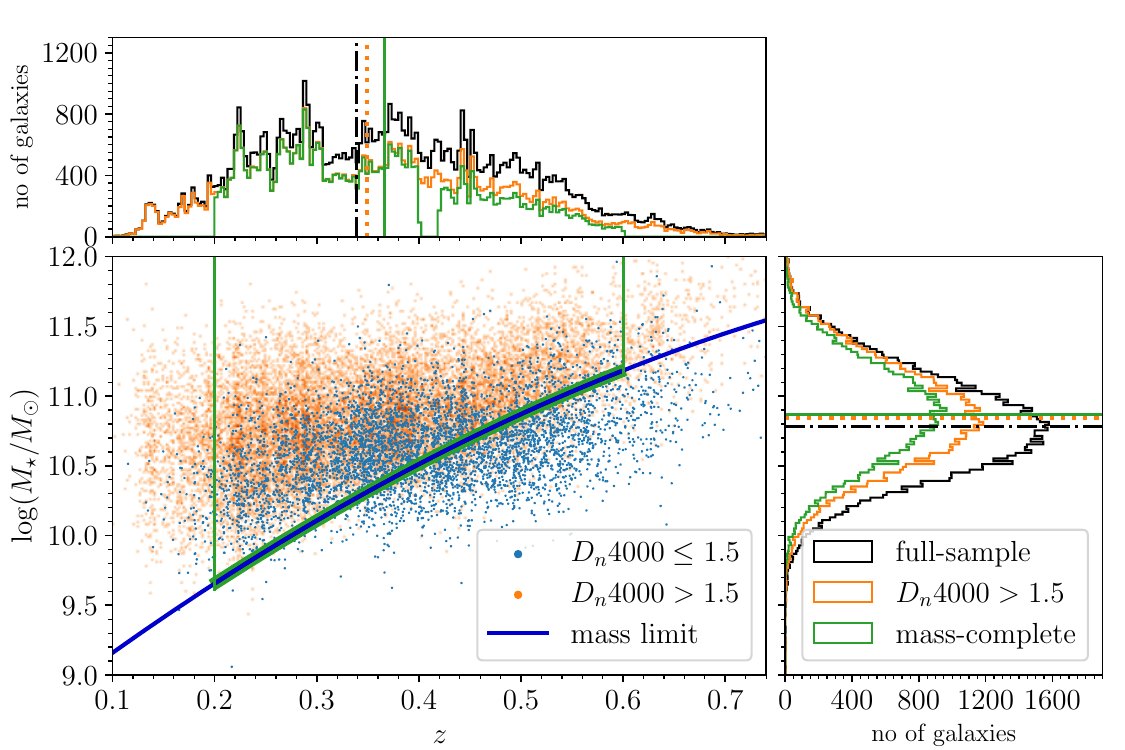}
    \caption{Stellar mass and redshift of HectoMAP galaxies. Blue and orange points in the central panel show $M_\star$ as a function of $z$ for the HectoMAP subsample of color-selected galaxies with measured $M_\star$. The orange points show HectoMAP quiescent galaxies ($D_n4000>1.5$). The blue line indicates the stellar mass limit as a function of redshift based on the survey magnitude limit. Quiescent (orange) galaxies within the region outlined by green solid lines are included in the mass-complete quiescent sample. The upper and right panels respectively show the marginal distributions in $z$ and $M_\star$ for the full (black), quiescent (orange), and mass-complete (green) HectoMAP samples. In these panels, the black dash-dotted, orange dotted, and green solid lines indicate medians of the correspondingly colored distributions. To improve readability in the central panel we show only a randomly selected $30\%$ of the total galaxy sample. Marginal distributions refer to the total samples.}\label{fig:masscomp}
\end{figure*}

Figure~\ref{fig:masscomp} shows galaxy stellar mass as a function of redshift for the parent sample of 72,110 color-selected HectoMAP galaxies (blue and orange points in the central panel; the fifth row of Table~\ref{table:hectomap}). The black histograms in the upper and right panels show the marginal distributions of these galaxies in $z$ and $M_\star$, respectively. The median redshift and stellar mass (black dash-dotted line in upper and right panel) are $z=0.34_{-0.07}^{+0.09}$ and $\log(M_\star/M_\odot)=10.78_{-0.24}^{+0.25}$, respectively. Sub- and superscripts show the interquartile ranges.

The HectoMAP data releases also include the spectral indicator $D_n{4000}$, a ratio of fluxes around the $4000$~\AA\ break. $D_n{4000}$ is a proxy for the average age of the galaxy stellar population  \citep{Kauffmann03,Damjanov24}. 
Following the definition from \citet{Balogh1999}, we compute the ratio of fluxes in two spectral windows: (3850–3950)~\AA\ and (4000–4100)~\AA\/. The median S/N at (3850–4100)~\AA\ is $\sim 4.5$. We measure $D_n4000$ for $\sim 96.5\%$ of the color-selected HectoMAP galaxies with mass measurement (sixth row of Table \ref{table:hectomap}).

Based on the estimated range of spectral indices from a sample of synthetic galaxy spectra \citep{Kauffmann03} and the signal-to-noise of HectoMAP spectra (Table~1 and Fig.~7 of \citealt{Damjanov23}), we evaluate the reliability of the $D_n{4000}$ measurements for the parent galaxy sample. The criteria we use are: (1) $D_n4000>0$, (2) $D_n4000<3$, and (3) $\Delta D_n4000< 0.2D_n4000$. 

For galaxies with $D_n{4000}<2.2$, the measurement error $\Delta D_n4000$ does not depend on $D_n{4000}$; in contrast, $\Delta D_n4000$ increases with $D_n{4000}$ for galaxies with $D_n{4000}>2.2$. 
We resample $D_n{4000}$ for galaxies with $2.2<D_n{4000}<3$ by first computing the average ${\Delta D_n4000}$ -- $\overline{\Delta D_n4000}$ -- for all of the galaxies in this subset. Then for each galaxy, $i$, from the subset we randomly generate a set of corrections $\varepsilon_{i_j}$, where $j\in [1,100]$, from a normal distribution with zero mean and dispersion equal to $\overline{\Delta D_n4000}$. Using the average $\bar{\varepsilon_i}$ of these corrections, we computed ${D_n4000}_i^{new}={D_n4000}_i + \bar{\varepsilon_i}$. We retained galaxy $i$ with its corresponding ${D_n4000}_i^{new}$ in the sample if this new value is smaller than 2.2. This correction procedure for $D_n{4000}$ applies to only $\sim 4.4\%$ of the galaxies with $D_n4000$ measurements;  we removed $\sim 50\%$ of galaxies with initial index values $2.2<D_n4000<3$. The average measurement error $\Delta D_n4000$ for the final sample is 0.067. In Sect. \ref{subsec:masscomplete}, we describe how we used $D_n{4000}$ to select quiescent HectoMAP galaxies.

\subsection{The mass-complete sample of HectoMAP quiescent galaxies}\label{subsec:masscomplete}

We explored the evolution in both average size and light concentration of massive quiescent galaxies, measured at fixed rest-frame wavelength in a mass-complete sample \citep{Damjanov23}. We also examined the dependence of size and light concentration on galaxy stellar mass. A mass-complete sample is critically important for tracing evolutionary trends. 

To identify the quiescent population in the HectoMAP sample, \citet{Damjanov23} selected galaxies with $D_n4000>1.5$. This index is a robust indicator of quiescence for several reasons. First, the narrow wavelength range of the $D_n4000$ measurement (Sect.~\ref{subsec:hectomap}) minimizes its dependence on dust. Second, the selection of quiescent samples based on $D_n4000$ is not overly sensitive to changes in stellar metallicity \citep[e.g.,][]{Kauffmann03, LeBorgne06}.  Using the 30-band photometry and the MMT/Hectospec spectroscopy of galaxies in the COSMOS field, \citet{Damjanov18} showed that the selection of quiescent samples at $z\lesssim0.5$ based on $D_n4000$ is essentially equivalent to selection based on the rest-frame UVJ colors derived from photometric measurements over a very broad UV-infrared wavelength range. These advantages make the cut in $D_n4000$ a preferred method for selecting quiescent galaxies in spectro-photometric studies at $z<1$ \citep[e.g.][]{Vergani08, Woods10, Geller14, Damjanov18, Damjanov19, Damjanov23, Figueira24}.

There are 51,274 quiescent galaxies in HectoMAP (seventh row of Table \ref{table:hectomap}).
The orange points in the central panel of Fig. \ref{fig:masscomp} show the $M_\star-z$ relation of HectoMAP quiescent galaxies. The orange histograms in the upper and right panels show the marginal distribution in $z$ and $M_\star$ for these galaxies. 

The HectoMAP quiescent sample is incomplete at $z< 0.2$ \citep{Sohn21HDR1,Sohn23}. As a consequence of the red selection of HectoMAP, the survey is missing a significant fraction of the low-mass quiescent galaxies that are bright enough to be included at low redshift. In \citet{Damjanov23} we use the Smithsonian Hectospec Lensing Survey Field 2 sample \citep[SHELS,][]{Geller14,Geller16} to build a mass-complete quiescent sample from the HectoMAP parent sample. Although the SHELS F2 survey is slightly shallower than HectoMAP (with a limiting $r-$band magnitude $r_{\rm max}^{\rm SHELS}=20.75$), SHELS F2 is not color-selected. We translate the apparent magnitude limits of the  HectoMAP and SHELS F2 samples into corresponding absolute magnitude limits as a function of redshift. Using the distribution of $M/L$ ratios for galaxies in narrow redshift bins, we then compute the redshift-dependent stellar mass limits for the two complete quiescent galaxy samples. The blue curve in Fig. \ref{fig:masscomp} shows the stellar mass limit for HectoMAP. 

Next, we used SHELS F2 for assessing the redshift interval where HectoMAP provides a complete mass-limited sample of quiescent galaxies. We compared the mass distributions of SHELS F2 quiescent galaxies and all SHELS F2 galaxies with the HectoMAP color selection in $\Delta z=0.1$ redshift bins spanning $0.1<z<0.6$. 
For galaxies in the $0.2<z<0.6$ redshift range, the red color-selected SHELS F2 samples include 90\%-100\% of galaxies in the quiescent SHELS F2 samples. However in the $0.1<z<0.2$ redshift bin the color-selected sample is significantly smaller than the complete quiescent sample.

After removing galaxies at $0.4<z<0.42$, where a prominent night sky line significantly impacts $D_n4000$ measurements \citep{Damjanov24}, we concluded that HectoMAP provides a complete mass-limited quiescent sample of 39,468 galaxies covering the redshift range $0.2<z<0.6$ (eighth and ninth entries in Table \ref{table:hectomap}). Next, we analyzed this quiescent sample.
In Fig. \ref{fig:masscomp}, the quiescent galaxies (orange points) enclosed in the region outlined by the green solid lines constitute the HectoMAP mass-compete quiescent sample. The green histograms in the upper and right panels show the marginalized distribution along the corresponding axis. The median redshift and stellar mass of the sample (green lines in the upper and right panels) are $z=0.37_{-0.09}^{+0.09}$ and $\log(M_\star/M_\odot)=10.87_{-0.24}^{+0.24}$, respectively. This large sample provides a powerful platform for examining the evolution of massive quiescent galaxy population. 

\section{Galaxy morphology from Subaru/HSC imaging}\label{sec:hsc}

We characterize variation in the structural parameters of quiescent galaxies as a function of rest-frame wavelength. We also explore the dependence of these variations on redshift and stellar population age.
We then describe the measurements of morphological parameters for the HectoMAP galaxies based on HSC-SSP multiband imaging from Subaru/HSC. We summarize the HSC-SSP imaging of the HectoMAP field in Sect. \ref{subsec:hscintro}; we describe the single-Sérsic model fitting procedure in Sect. \ref{subsec:sepp}; finally, we report the results in Sect. \ref{subsec:ress}.

\subsection{HSC imaging}\label{subsec:hscintro}

The HSC-SSP PDR3 \citep{Aihara22} provides $g\text{-,}$ $r\text{-,}$ $i\text{-,}$ $z\text{-,}$ and $y\text{-}$band imaging of the HectoMAP region. The survey imaging covers the full HectoMAP field in the $g$, $r$, and $z$bands, and 41~deg$^2$ ($\sim$80\% of the survey) in the $i$ and $y$ bands. 

The image quality of HSC-SSP is excellent.
The median seeing (i.e., the full width at half maximum (FWHM) of the PSF) is $0\farcs 83$, $0\farcs 74$, $0\farcs 55$, $0\farcs 62$, and $0\farcs 69$ for the $g$, $r$, $i$, $z$, and $y$ band, respectively. In addition to the wavelength dependence of atmospheric seeing, different visiting times for different bands and changes in observing conditions also contribute to variations in image quality. The seeing is remarkably stable across the survey area: in each band, the local seeing is generally within $\sim 0\farcs 2$ of the median value\footnote{For HSC-SSP quality assurance plots, see https://hsc-release.mtk.nao.ac.jp/doc/index.php/quality-assurance-plots\_\_pdr3/.}.

The hscPipe system \citep{Bosch18} is the standard pipeline for HSC-SSP \citep{Miyazaki18}. The pipeline facilitates reduction of individual chips, mosaicking, and image stacking. The HSC-SSP DR3 offers two sets of coadded images based on two different approaches to sky subtraction: local and global \citep[see Fig. 8 of][]{Aihara22}. 
We select coadds with local subtraction to mitigate the contribution of the extended wings of bright stars that may contaminate background galaxies. Only bright sources lose their wings with local sky subtraction. Sources with an extent $\lesssim 30$~arcsec ($\sim 100-200$~kpc at $0.2<z<0.6$) remain unaffected \citep{Aihara22}. The sample we analyze does not contain galaxies that approach this size limit (see Sect. \ref{subsec:ress}). 

\subsection{Single-Sérsic model fitting pipeline}\label{subsec:sepp}

We obtain morphological parameters by fitting the 2D galaxy surface brightness in each of the five HSC bands with the Sérsic intensity model \citep{Sersic68}. The 1D radial dependence of the model is
\begin{equation}\label{eq:sersic}
    I(r) = I_0 \exp{\left\{ -b(n) \left(\frac{r}{R_e} \right)^{\frac{1}{n}} \right\} },
\end{equation}
where $I_0$ is the central intensity (surface brightness), $R_e$ is the half-light radius along the galaxy major axis, and $n$ is the Sérsic index, which describes the concentration of the model profile. The coefficient $b(n)$, a function of $n$,  ensures that $R_e$ encloses half of the total galaxy light \citep{Ciotti99}. 

Extraction of Sérsic model parameters involves three steps. First, we run the \se software \citep{Bertin96} to obtain a catalog that includes detection of point-like sources. Then, we run the \texttt{PSFex} software \citep{Bertin11} on this catalog to construct a set of spatially varying PSFs. The PSF model at the position of an extended source is a linear combination of basic vectors that best fits the observed point-source profiles in the vicinity of that position. Finally, we run the \sepp software \citep{Bertin20,Kummel20}\footnote{Comprehensive \sepp documentation is in \texttt{https://sextractor.readthedocs.io/en/latest/index.html}.} to fit  the galaxy surface brightness profile with a PSF-convolved Sérsic model (Eq. (\ref{eq:sersic})) in 2D. Following this procedure, we built the catalog of morphological parameters for the HectoMAP galaxy sample.

\sepp, the successor to the original \se package, fits models to sources by minimizing a least-squares loss function with respect to a parameter vector describing the shape of the 2D Sérsic model. \sepp includes a flexible and Python-based model configuration, the possibility of defining complex priors and dependences for model parameters, and code optimization for fast processing.

\citet{Bretonniere23} analyzed the performance of the \sepp model fitting by using realistic simulations of galaxies in \textit{Euclid}'s visible band $I_E$. This broadband range covers $\approx (5400-9000)~$~\AA\ \citep{Laureijs11}, overlapping with the $r$, $i$, and $z$ bands of HSC. For single-Sérsic model fitting to realistic synthetic profiles, the software provides reliable structural measurements with a bias $\lesssim 15\%$ and a scatter $\lesssim 30\%$ for apparent magnitude  $I_E\sim23$. Model uncertainties are typically underestimated by a factor $\sim 2$. \citet{Bretonniere23} test \sepp along with other source-extraction and fitting software packages developed for large-area imaging surveys and conclude that \sepp is the best approach.

The \sepp 2D Sérsic model fitting provides best-fit values for seven free parameters: the 2D cartesian coordinates of the galaxy center, the ellipse rotation angle $\theta$, the ellipse minor-axis to major-axis ratio $b/a$, along with $I_0$, $R_e$, and $n$. Fits to the spatial model parameters are given in pixel units, which are then converted to arcseconds. We applied a standard initialization for the seven fitting parameters:
\begin{itemize}
    \item the \sepp isophotal centroid for the 2D cartesian coordinates of the galaxy center, allowing it to vary linearly in the range limited to the isophotal half-light radius (from the built-in function {\texttt{get\_pos\_parameters}});
    \item isophotal proxy for $\theta$, spanning the range $(-\pi,+\pi)$ linearly;
    \item $b/a=0.5$, allowing  for exponential variation  within the  range $(0.1,1)$;
    \item isophotal guess for the central surface brightness $I_0$, allowing for exponential variation within $\pm 10^3$ times the initial proxy (from the built-in function \texttt{get\_flux\_parameter});
    \item isophotal half-light radius for $R_e$,  for  exponentially spaced bins within  $(0.1,10)\times$ the isophotal value interval;
    \item $n=1$, allowing for linear variation in the range $(0,10)$.
\end{itemize}
We also considered a mild non-Gaussian prior for the Sérsic parameter, $n$, to minimize the chance of obtaining (large) final values of $n\sim 9-10$. We require  the random variable $N=\exp{\left(n-n_0\right)}$, with $n_0=8$, to be normally distributed with zero mean and standard deviation equal to 1. This prior is constant (uninformative) for $n\lesssim 7$, and decreases for larger $n$: it assigns a starting probability of $\sim 1/2$ to $n\sim8 $ and a probability of $\sim 0$ to $n\gtrsim 9$.

\subsection{Sérsic fitting results}\label{subsec:ress}

For each of the five HSC photometric bands, we measured the Sérsic morphological parameters for all of the galaxies in the HectoMAP sample with HSC-SSP PDR3 imaging. We obtained best-fit models for all HectoMAP targets in the region with images available in all five bands ($\sim 78,000$ galaxies, or 82\% of the parent HectoMAP sample; Sect. \ref{subsec:hscintro}, second row in Table \ref{table:hectomap}).

\sepp identifies HectoMAP galaxies in HCS-SSP by requiring that the fitted model centroid lie within a radius of five pixels from the input centroid listed in the HectoMAP datatbase. The angular distance between the model and input centroid is $< 50\%$ of the best ($i$ band) PSF FWHM for $>99\%$ of measured galaxies (Sect. \ref{subsec:hscintro}).
After identifying all of the targets in the HectoMAP catalog, \sepp proceeds to fit a 2D Sérsic model to each galaxy surface brightness profile (Sect. \ref{subsec:sepp}). Table \ref{table:measures} lists the resulting best-fit parameters $R_e$, $b/a$, and $n$, along with $z$ and $M_\star$.

\begin{table*}[h]
\begin{center}
\caption{\label{table:measures} Single-Sérsic model structural parameters for HectoMAP galaxies.$^\star$}
\begin{tabular}{ccccccc}
\hline
\hline
 ID & \begin{tabular}{c} Position \\ (RA+DEC) \end{tabular} & $z_{\rm spec}$ & $\log(M_\star/M_{\odot})$ & $R_e ('')$ & $b/a$ & $n$\\
\hline
123$\dots$231 & \begin{tabular}{l} 249.99821+\\42.97923 \end{tabular} & 0.2245$\pm$0.0003   & 9.3$\pm$0.5  &  \begin{tabular}{r} 
0.85 $\pm$ 0.04 $(g)$ \\ 
0.86$\pm$0.02 $(r)$ \\ 
0.83 $\pm$  0.01 $(i)$  \\ 
0.80 $\pm$ 0.01 $(z)$ \\ 
0.82 $\pm$ 0.03 $(y)$  
\end{tabular}  & \begin{tabular}{r} 
0.30 $\pm$ 0.03 $(g)$ \\  
0.33$\pm$0.02 $(r)$ \\ 
0.318 $\pm$     0.007 $(i)$ \\ 
0.310 $\pm$ 0.005 $(z)$ \\ 
0.32 $\pm$ 0.02 $(y)$ 
\end{tabular}  & \begin{tabular}{r}     
1.3$\pm$0.1 $(g)$ \\ 
1.3 $\pm$ 0.2 $(r)$ \\ 
1.31 $\pm$ 0.05 $(i)$ \\ 
1.26 $\pm$ 0.04 $(z)$ \\ 
1.7 $\pm$ 0.2 $(y)$ 
\end{tabular}   \\
\hline                                  
123$\dots$559 & \begin{tabular}{l}249.98666+\\ 43.06497\end{tabular} & 0.3685$\pm$0.0002  & 10.3$\pm$0.2  & \begin{tabular}{r}
0.64 $\pm$ 0.02 $(g)$ \\
0.645 $\pm$ 0.008 $(r)$ \\
0.570 $\pm$ 0.005 $(i)$ \\
0.55 $\pm$ 0.01 $(z)$ \\
0.51 $\pm$ 0.02 $(y)$
\end{tabular} & \begin{tabular}{r}
0.40 $\pm$ 0.02 $(g)$ \\
0.380 $\pm$ 0.008 $(r)$ \\
0.413 $\pm$ 0.005 $(i)$ \\
0.435 $\pm$ 0.007 $(z)$ \\
0.40 $\pm$ 0.02 $(y)$
\end{tabular} & \begin{tabular}{r}
2.00 $\pm$ 0.09 $(g)$ \\
2.46 $\pm$ 0.05 $(r)$ \\
2.47 $\pm$ 0.05 $(i)$ \\
2.68 $\pm$ 0.08 $(z)$ \\
3.01 $\pm$ 0.28 $(y)$
\end{tabular} \\
\hline                                                                  
123$\dots$773 & \begin{tabular}{l}249.98529+\\ 43.27685\end{tabular} & 0.3527$\pm$0.0006  & 11.1$\pm$0.2  & \begin{tabular}{r}
0.88 $\pm$ 0.02 $(g)$ \\
0.85 $\pm$ 0.01 $(r)$ \\
0.731 $\pm$ 0.007 $(i)$ \\
0.578 $\pm$ 0.005 $(z)$ \\
0.67 $\pm$ 0.01 $(y)$
\end{tabular} & \begin{tabular}{r}
0.93 $\pm$ 0.02 $(g)$ \\
0.904 $\pm$ 0.009 $(r)$ \\
0.848 $\pm$ 0.005 $(i)$ \\
0.749 $\pm$ 0.005 $(z)$ \\
0.76 $\pm$ 0.01 $(y)$
\end{tabular} & \begin{tabular}{r}
3.3 $\pm$ 0.1 $(g)$ \\
3.35 $\pm$ 0.08 $(r)$ \\
3.18 $\pm$ 0.05 $(i)$ \\
2.13 $\pm$ 0.03 $(z)$ \\
3.8 $\pm$ 0.1 $(y)$
\end{tabular} \\
\hline
$\cdots$ & $\cdots$ & $\cdots$ & $\cdots$ & $\cdots$ & $\cdots$ & $\cdots$ \\
 \hline
 \end{tabular}
 \end{center}
 {{\bf Notes.} This table is available in its entirety at the CDS. A portion is shown here for guidance regarding its form and content.}
 \end{table*}

There are 28,128 galaxies in the HectoMAP mass-complete sample of quiescent galaxies with measured structural parameters in all five photometric bands (Sect. \ref{subsec:masscomplete}, next-to-last entry of Table \ref{table:hectomap}). A small fraction (4.3\%) of these galaxies have measured $R_e$ smaller than half of the typical seeing in one or more bands. We remove these essentially unresolved galaxies from the sample. The final sample with reliable single S\'ersic profile parameters in all five HSC band includes 26,912 quiescent galaxies (last entry of Table \ref{table:hectomap}).

The median reduced $\chi^2$ of the fits is in the range $(1.3-6.7)$, with the 25th, 50th, and 75th percentile of the reduced $\chi^2$ of $(1.1,1.5,3.0)$, $(1.4,2.8,9.1)$, $(2.4,6.7,26.0)$, $(1.3,2.5,8.9)$, and $(1.1,1.3,2.6)$ for the HSC $g$, $r$, $i$, $z$, and $y$ band, respectively. The average goodness-of-fit magnitude $\chi^2\sim1$ confirms small deviations between the best-fit single-S\'ersic models and the light profiles of quiescent HectoMAP galaxies.
In Appendix \ref{app:comparison} we further compare the $i-$band fits with independent estimates of morphological parameters derived by fitting the HSC $i-$band images of the HectoMAP quiescent sample using \se \cite{Damjanov23}. The agreement is excellent between the two sets of measurements.

\begin{figure*}[h]
    \centering
    \includegraphics[width=\linewidth]{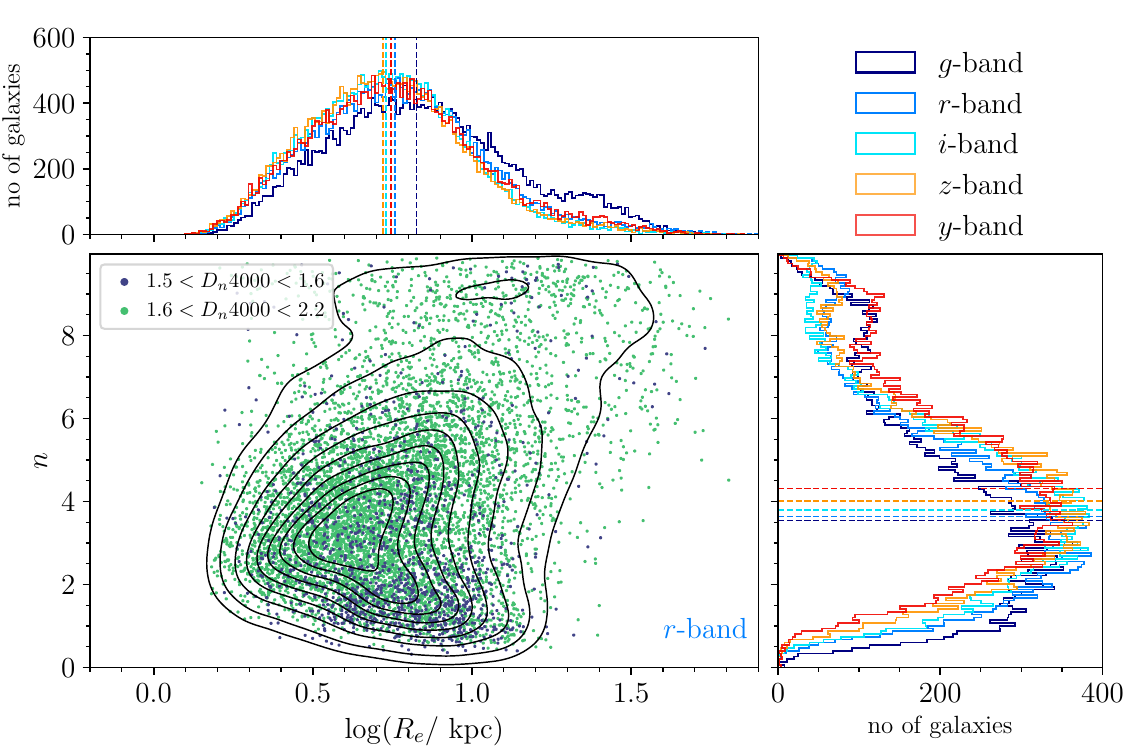}
    \caption{Half-light radius, $R_e$, and Sérsic index, $n$, in five HSC bands for 26,912 quiescent galaxies. These galaxies have morphological measurements in all five bands and $R_e$ exceeds half of the seeing FWHM. The central panel shows the distributions of $r\text{-}$ band in the $R_e-n$ plane. Galaxies (points) are segregated into two classes of $D_n4000$: $1.5<D_n4000<1.6$ (dark blue) and $D_n4000>1.6$ (green). Black contours show the probability density of HectoMAP galaxies in the $R_e-n$ plane derived from a Gaussian kernel estimator (KDE). The largest to the smallest contour enclose regions where the probability density exceeds the 10th to the 90th percentiles of the distribution. The upper and right panels show the marginal distributions in $R_e$ and $n$, respectively, color-coded by the HSC band. The equivalently color-coded dashed lines indicate the medians of these distributions. To improve readability, the central panel displays a randomly selected $30\%$ of the total sample. Histograms and isocurves are based on the total samples.}
    \label{fig:distr_RN_all}
\end{figure*}

The large panel in Fig. \ref{fig:distr_RN_all} shows the distribution of measurements in the $R_e-n$ plane for the $r\text{-}$band. Results for the other bands are qualitatively similar. For spheroidal galaxies with $n\gtrsim 2$ ($\sim 85-95\%$ of the sample) the expected positive correlation between galaxy size and concentration is evident \citep[e.g.,][]{Caon1993,Trujillo01,Schombert13,Vika15}. Less centrally concentrated galaxies (with $n\sim 1$) have larger radii along the semi-major axis ($R_e$) on average. There is an anticorrelation between size and concentration. 

Nearly all spheroidal galaxies ($n>2$) have old stellar populations ($D_n4000>1.6$, green points in Fig. \ref{fig:distr_RN_all}). Quiescent galaxies with younger stellar age ($1.5<D_n4000<1.6$, blue points in the figure) populate the small-$n$ and large-$R_e$ (bottom right) region of the $R_e-n$ plane. These galaxies, newcomers to the quiescent population (Sects. \ref{subsec:masscomplete} and \ref{subsec:grad_dn4000}), drive the anticorrelation between galaxy size and Sérsic index.

A comparison of average size and Sérsic index for quiescent galaxies with older and younger stellar ages elucidates the difference in structural properties of these two subpopulations. For example, in the $r\text{-}$band, galaxies with $1.5<D_n4000<1.6$ have on average $\log (R_e/$~kpc$)\sim 0.79$ and $n\sim 2.4$; galaxies with older stellar populations ($D_n4000>1.6$) have $\log (R_e/$~kpc$)\sim 0.75$ and $n\sim 3.9$. We observe qualitatively equivalent differences in the remaining four bands. Newcomers to the quiescent population are consistently larger and less concentrated than quiescent galaxies with older stellar population age.

The histograms in the upper and right panels of Fig. \ref{fig:distr_RN_all} show the distributions of $R_e$ and $n$ for the final sample of quiescent galaxies, color-coded by the HSC band.
The distributions show similar shapes across all five bands. Sizes generally decrease as the wavelength increases: the median $\log (R_e/$~kpc$)$ is $\sim 0.83,0.76,0.73,0.72,$ and $0.75$ in the $g$, $r$, $i$, $z$, and $y$ band, respectively. In contrast, Sérsic indices increase with increasing wavelength: the median $n$ is $\sim 3.6,3.7,3.8,4.0,$ and $4.3$ from the $g\text{}$ to the $y\text{ }$band.

A small fraction ($\sim 9.3\%$) of galaxies in our quiescent sample have $n>8$ in all bands (the secondary peaks in the right panel of Fig. \ref{fig:distr_RN_all}).  On average, galaxies with $n>8$ are $\sim 0.26$~mag fainter than galaxies with $n<8$.  The fit residuals for $n>8$ galaxies show no significant difference with respect to the rest of the population. Moreover, the average reduced $\chi^2$ for galaxies with $n>8$ is within the interquartile range of the $\chi^2$ distribution for the subsample with $n<8$. We perform the analysis from Sects.~\ref{sect:results} and~\ref{sect:discussion} with and without high-$n$ galaxies and obtain indistinguishable results. To provide a comprehensive analysis, we report the results using the complete mass-limited sample.

The upper panel in Fig. \ref{fig:distr_RN_all} shows a small secondary peak in the size distribution for the $g$ band, at $\log (R_e/$~kpc$)\sim 1.2-1.6$. Only $\sim 100$ galaxies are responsible for this feature. The $g$ band has the largest median seeing, $\sim 0.83$~arcsec, thus this lower image quality could cause overestimation of $R_e$ and $n$ for the most centrally concentrated galaxies. However, because this issue affects only $\lesssim 0.5\%$ of the sample, the overall results are unaffected.

\section{Multiwavelength Sérsic profiles}\label{sect:results}

The HectoMAP data set enables the investigation of trends in the rest-frame structural parameters as a function of wavelength for galaxies segregated in redshift, stellar mass, and average stellar population age. Section \ref{subsec:grad} outlines the procedure for measuring trends with rest-frame wavelength and the derived dependence on $z$ and $M_\star$. Section \ref{subsec:grad_dn4000} investigates the changes in the wavelength dependence of S\'ersic structural parameters (i.e. ,$R_e$ and $n$) as a function of $D_n4000$.

\subsection{Variations in galaxy size and Sérsic index with wavelength} \label{subsec:grad}

In the redshift range  $0.2<z<0.6$, the rest-frame wavelengths ($\lambda^{RF}$) are up to $\sim 38\%$ smaller than the observed wavelengths ($\lambda_f$) in the five HSC bands ($f$). At $z=0.6$, the photometric bands correspond to rest-frame wavelengths that are  $\sim 25\%$ bluer $\lambda^{RF}$ than at $z=0.2$. Light profiles covering a range of $\lambda^{RF}$ enable examinations of the distributions of stellar populations with different ages within a galaxy. Generally the morphological parameters depend on $\lambda^{RF}$. 
The range of wavelengths spanned by the five bands of HSC, $~(4,000-10,700)~$~\AA\ (Sect. \ref{subsec:hscintro}) along with precise redshift determinations from Hectospec spectra enable the determination of morphological parameters as a function of rest-frame wavelength.

We define the rate of change in the Sérsic model parameter $\mathcal{P}$ with wavelength as 
\begin{equation}\label{eq:gradR}
    \frac{\Delta \log \mathcal{P} }{\Delta \log \lambda^{RF}}=\frac{\log(\mathcal{P}^{RF7000})-\log(\mathcal{P}^{RF3500})}{\log(7000~{[\mathring{A}}])-\log(3500{[\mathring{A}}])}.
\end{equation}
To obtain sizes, we set $\mathcal{P} = R_{e,c}[$kpc$]$, where $R_{e,c}$ is the circularized half-light radius. This radius averages out the galaxy asphericity; it is defined as $R_{e,c}=R_e \sqrt{b/a}$, where $b/a$ is the axis ratio. For Sérsic index, $\mathcal{P}=n$. 

In Eq. (\ref{eq:gradR}), we consider the differences between two rest-frame parameters, $\mathcal{P}^{RF}$, at wavelengths of $\lambda^{RF}_1=3500~$~\AA\ and $\lambda^{RF}_2=7000~$~\AA\/. This choice enables the exploration of the full dynamic range of observed wavelengths spanned by the HSC photometry without extrapolating beyond the observed range.

Based on Eq. (\ref{eq:gradR}), we first estimate $\mathcal{P}^{RF}$ at $\lambda^{RF}_1$ and $\lambda^{RF}_2$ for all  26,912 galaxies in the five bands (Table \ref{table:hectomap}). 
For each galaxy, we determine the best linear fit to the five bands $(\lambda_f, \mathcal{P}_f)$,
\begin{equation}\label{eq:linfit}
    \log(\mathcal{P}_f) = a_{\mathcal{P}} \lambda_f + b_{\mathcal{P}}.
\end{equation}
The variable $\lambda_f$ includes the central observed wavelength of the five HSC bands: $\lambda_f=4750, 6225, 7700, 8875$, and $10100~$~\AA\ for the $g$, $r$, $i$, $z$, and $y$ bands, respectively. Using HectoMAP spectroscopic redshifts we evaluate Eq. (\ref{eq:linfit}) at $\lambda_{1,2} = (1+z)\lambda^{RF}_{1,2} $ to estimate $R_{e,c}^{RF}$ and $n^{RF}$ in the two required rest frame wavelengths.

We use Eq. (\ref{eq:gradR}) to determine variations in $R_{e,c}$ and $n$ with wavelength. We examine the data in four equally spaced bins of redshift and a variable number of equally spaced $\Delta\left[\log(M_{\star}/M_\odot)\right]=0.25$~dex bins in stellar mass. In Table \ref{table:bin_size} we report the total number of galaxies in our four redshift bins and the corresponding range of lower limits on $\log (M_\star/M_\odot)$ for the mass-complete sample (blue line in the central panel of Fig. \ref{fig:masscomp}). The total number of galaxies decreases with increasing redshift. However, the sample size remains statistically large in all four redshift bins.
\begin{table}[htbp]
\begin{center}
\caption{\label{table:bin_size} Sampling and mass limit for the HectoMAP redshift bins.}
\begin{tabular}{lcc}
\hline
\hline
$z$~range & no. of galaxies & $\log(M_{\ast,lim}/M_\odot)$ \\
\hline
$0.2<z<0.3$ & 10863 & (9.66-10.11) \\
$0.3<z<0.4$ & 8439 & (10.11-10.51) \\
$0.4<z<0.5$ & 4720 & (10.59-10.87) \\
$0.5<z<0.6$ & 2890 & (10.87-11.18) \\
\hline
\end{tabular}
\end{center}
{{\bf Notes.} The lower and upper limit of the intervals in $\log(M_{\star,lim}/M_\odot)$ correspond to the values of the $M_{\star,lim}(z)$ function (blue curve in Fig. \ref{fig:masscomp}) at the lower and upper edges of the four redshift bins from the first column.}
\end{table}

The lower and upper masses sampled in our mass bins are $\log (M_\star, min/M_\odot)=10.0$ and $\log (M_\star, max/M_\odot)=11.5$, respectively. This choice enables us to include a minimum of 300 galaxies in each equally spaced $\Delta\left[\log(M_{\star}/M_\odot)\right]=0.25$~dex bin in stellar mass at any redshift. The number of stellar mass bins decreases with increasing redshift because the lower $M_{\star}$ limit increases with redshift. As a result, the number of mass bins decreases from eight to two from $0.2<z<0.3$ to $0.5<z< 0.6$. At $0.2<z<0.3$, the six bins of increasing mass contain 996, 2432, 3244, 2660, 1212, and 319 galaxies, respectively. At $0.5<z< 0.6$, the two mass bins contain 1645 and 1245 galaxies. Overall, the mass bins contain a statistically large number of galaxies regardless of redshift and corresponding mass limit. 

\begin{figure}
    \centering
    \includegraphics[width=\linewidth]{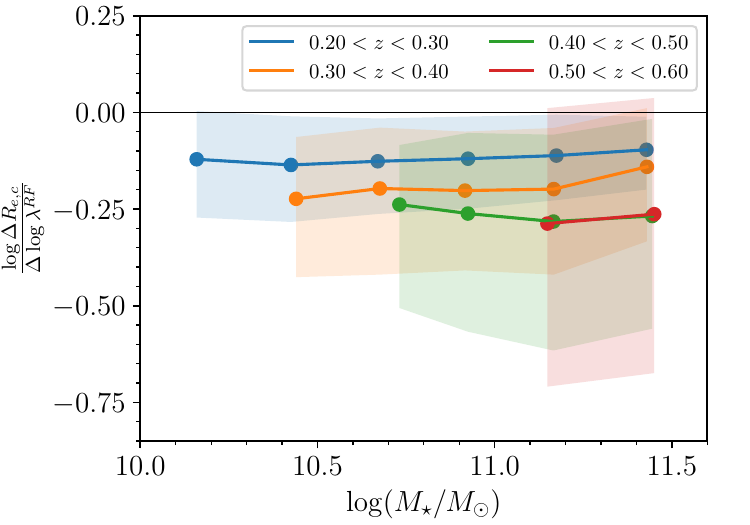}
    \caption{Rate of change in $R_{e,c}$ with rest-frame wavelength. Blue, orange, green, and red circles show the median $\frac{\Delta \log R_{e,c}}{\Delta \log \lambda^{RF}}$ in four bins of increasing redshift listed in the legend. Shadowed bands in matching colors indicate the interquartile ranges of the distributions. }
    \label{fig:gradR}
\end{figure}
Circles in Fig. \ref{fig:gradR} show the median rate of change in sizes $R_{e,c}$ with rest-frame wavelength. Size always decreases with  increasing wavelength, in the range  $\sim -(0.1-0.3)$;  $R_{e,c}$ decreases with increasing $\lambda^{RF}$. In other words,  as is well-known from other studies, the cores of quiescent galaxies are redder than their outer regions.

The absolute value of the size changes with wavelength at a rate that depends mildly on redshift. The median $\frac{\Delta \log R_{e,c}}{\Delta \log \lambda^{RF}}$ is $\sim -0.3$ for $z>0.4$, and $\sim -0.1$ for $0.2<z<0.3$, implying an average $\sim 30\%$ decrease in absolute value as the redshift decreases from $z\sim 0.4$ to $z\sim 0.2$. 
The interquartile range (shadowed areas in Fig.~\ref{fig:gradR}) shows that the relative scatter is $\sim 100\%$ at any redshift. These changes in size are independent of $M_\star$.

The HectoMAP-based estimates of $\frac{\Delta \log R_{e,c} }{\Delta \log \lambda^{RF}}$ at $0.2<z<0.4$ are within the range of values based on statistical spectroscopic samples at $z\lesssim0.25$ \citep[$\sim -0.16$, GAMA survey,][]{Kelvin12,Kennedy16} and much smaller spectroscopic samples from a broader redshift interval \citep[$-0.25$ for $0<z<2$, 3D-HST+CANDELS,][]{vdWel14}. A recent photometric study exploring the redshift range of HectoMAP also reports similar results for $\sim10^{11}\, M_\sun$ quiescent systems, with $\frac{\Delta \log R_{e,c} }{\Delta \log \lambda^{RF}}\sim -(0.10-0.23)$ at $z\sim 0.3$, and $\sim -(0.29-0.40)$ at $z\sim 0.6$ \citep[Table~3 from][]{George24}.

\begin{figure}
    \centering
    \includegraphics[width=\linewidth]{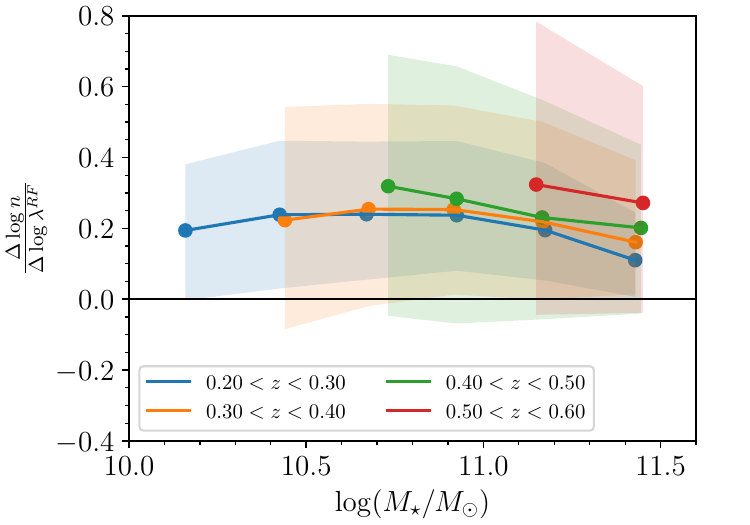}
    \caption{Same as Fig. \ref{fig:gradR} but for the Sérsic index.}
    \label{fig:gradn}
\end{figure}

The median rate of change in S\'ersic index with stellar mass and redshift (Fig. \ref{fig:gradn}) shows that the light profiles of quiescent galaxies are always more centrally concentrated at longer rest-frame wavelengths (i.e., $\frac{\Delta\log n}{\Delta \log\lambda^{RF}}$ is always positive). The rate of change in $n^{RF}$ with rest-frame wavelength is essentially independent of redshift or $M_\star$. The average $\frac{\Delta\log n}{\Delta \log\lambda^{RF}}$ is $\sim 0.20$ for $0.2<z<0.5$, and $\sim 0.3$ for $0.5<z<0.6$. The relative dispersions are $\sim 100\%$ independent of redshift and $M_\star$.

The sizes of quiescent galaxies at $0.2<z<0.6$ are larger at bluer wavelengths (Fig.~\ref{fig:gradR}). In contrast, their median Sérsic index increases with wavelength (Fig.~\ref{fig:gradn}). The red light profiles that trace the bulk of the galaxy stellar mass distribution are smaller and more centrally concentrated than the blue light profiles dominated by younger or more metal-poor stellar populations. The wavelength-dependent properties of light profiles in the HectoMAP sample confirm the results from spatially resolved photometric studies based either on statistically large samples of quiescent systems at $z\lesssim 0.2$ \citep[e.g.,][]{Peletier90,Tamura2000,LaBarbera05,LaBarbera09,Gonzalez11} or significantly smaller samples at $z\lesssim 2$ \citep[e.g.,][]{Guo11,Gargiulo12,Tacchella15}.

The fact that the cores of quiescent systems are redder than their outer regions has the potential to constrain quenching within galaxies \citep[i.e, inside-out quenching, where star formation terminates first in the central regions,][]{Avila18, NelsonE21, McDonough23, Mun24, Mun25}. However, galaxies grow in size during the quiescent phase and this growth is also wavelength-dependent \citep[e.g.,][Sect.~\ref{subsec:scalingrel}]{Naab09,Hilz2012,Oser2012}. To distinguish between the effects of the two processes, it is critical to compare the wavelength dependence of structural parameters for galaxies with different average stellar population ages. 

\subsection{Wavelength dependence of galaxy structural parameters across stellar population age (D$_n4000$)}\label{subsec:grad_dn4000}

The HectoMAP data trace, for the first time, changes in the  wavelength dependence of galaxy morphological parameters with galaxy average age based on the spectral index $D_n4000$. To explore the most extreme effects of this dependence, we compare structural parameters as functions of wavelength for a sample of newcomers to the quiescent population with a sample of aging galaxies that are already quiescent at $z\sim 0.6$. 

Newcomers are galaxies that recently joined the quiescent population. We identify newcomers at each redshift as galaxies with $1.5< D_n4000 < 1.6$, a standard selection in spectroscopic studies at $z\lesssim 1$ \citep{Kauffmann03,Vergani08,Damjanov23,Figueira24}. There are 3888 newcomers in the mass-complete sample.
Aging galaxies have $D_n4000>1.5$ at $z\sim 0.6$. At lower redshift, aging galaxies have larger $D_n4000$ values because of the relation between this spectral index and the galaxy global stellar population age.

In \citet{Damjanov23}, we developed a procedure for selecting the aging population in the HectoMAP sample based on $D_n4000$. Using the Flexible Stellar Population Synthesis code \citep[FSPS,][]{Conroy09,Conroy10}, we derived evolving synthetic spectra for galaxies that by $z=0.6$ have not been forming stars for 0, 1, 2, 3, or 4~Gyr. Redshift evolution from $z\sim0.6$ to $z\sim0.2$ of these synthetic spectra is based on the increase in  $D_n4000$ as the universe ages over this redshift range. The general increase in $D_n4000$ with decreasing redshift enable the selection of $0.2<z<0.6$ HectoMAP galaxies with average stellar population ages equivalent to the descendants of quiescent $z\sim0.6$ galaxies. There are 13,564 aging quiescent galaxies in our mass-complete sample. 

\begin{figure*}[htbp]
    \centering
    \includegraphics[width=\linewidth]{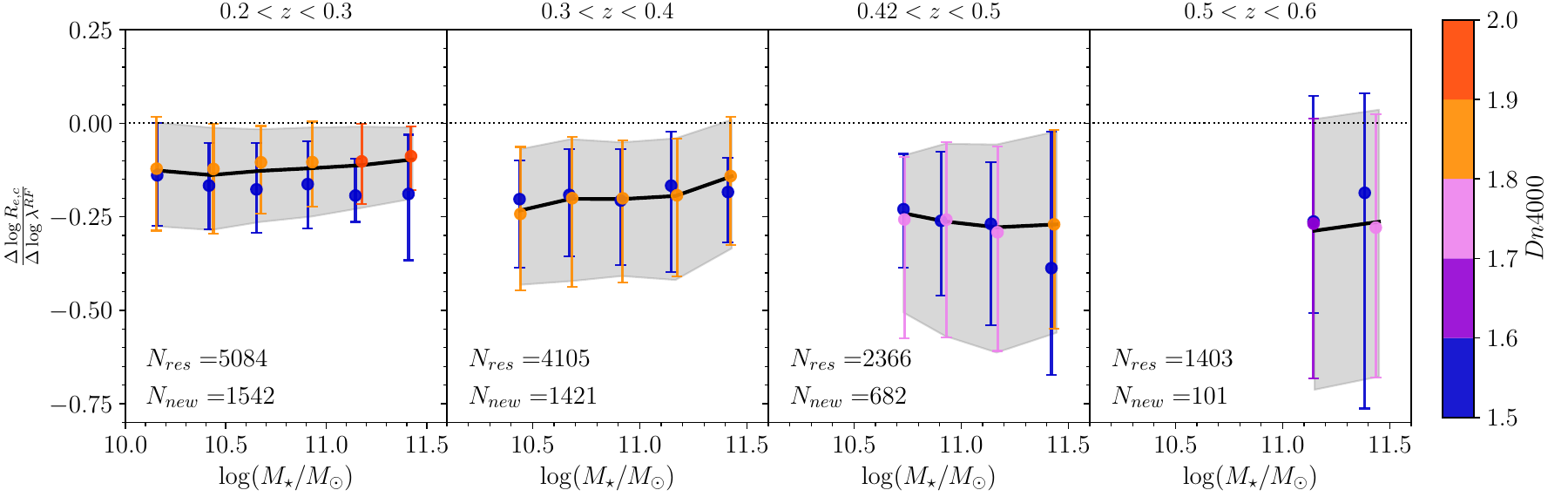}
    \caption{$\frac{\Delta \log R_{e,c}}{\Delta \log \lambda^{RF}}$ as a function of stellar mass and $D_n4000$ for two quiescent subpopulations. Circles show the median rate of change in size with rest-frame wavelength for the subpopulations of aging galaxies (shades of purple and red) and newcomers (blue) in $\Delta\left[\log(M_{\star}/M_\odot)\right]=0.25$ bins of stellar mass. The color of the symbols corresponds to the median $D_n4000$ index of the subpopulation. The panels show the four $\Delta z=0.1$ intermediate redshift bins (as indicated in the panel legends). Error bars denote the interquartile range of the correspondingly colored population. Colors indicate $D_n4000$. The black curve in each panel shows the median rate of change in size with rest-frame wavelength of the total quiescent population (same as Fig. \ref{fig:gradR}) and the gray area shows the corresponding interquartile range. }\label{fig:gradRdn4000}
\end{figure*}

Figure \ref{fig:gradRdn4000} illustrates the dependence of variations in galaxy size with rest-frame wavelength on the median $D_n4000$ index that separates newcomers (blue circles) from the aging quiescent population (purple- and red-shaded circles) in four $\Delta z=0.1$ redshift bins with $0.2<z<0.6$. In each redshift bin, the subpopulations are separated into $\Delta\left[\log(M_{\star}/M_\odot)\right]=0.25$ stellar mass bins.

For all stellar mass bins covering  $0.2<z<0.3$, $\frac{\Delta \log R_{e,c}}{\Delta \log \lambda^{RF}}$ for newcomers is $\sim 10-50\%$ larger in absolute value than for the aging population. The difference between the two subpopulations increases steadily with increasing $M_\star$. There is no difference between the variations in size with rest-frame wavelength for newcomers and the aging population for  $z>0.3$.

We use the Kolmogorov-Smirnov (KS) test to investigate the statistical significance of the difference between the wavelength-dependence of size for newcomers and the aging population. At $0.2<z<0.3$, the KS test yields a $p$-value of $\sim 10^{-3}-10^{-9}$ in all stellar mass bins except in the lowest and the highest bins. At $z>0.3$, the KS test cannot reject the null hypothesis that the decreases in size with wavelength for the two populations are drawn from the same distribution.  The decrease in subsample size and smaller stellar population age (i.e., D$_n 4000$) differences between the two quiescent populations limit the power of statistical tests at higher redshift. Nevertheless, the KS tests confirm that the difference between the decrease in size with wavelength for newcomers and aging galaxies, visible in the first panel of Fig.~\ref{fig:gradRdn4000}, is statistically significant.

At $0.2<z<0.3$ newcomers have a systematically smaller ratio between their sizes at redder and bluer wavelengths compared to aging galaxies. Relative to their stellar mass distribution, newcomers at this redshift have more extended blue light profiles than the aging subpopulation.

\begin{figure*}[htbp]
    \centering
    \includegraphics[width=\linewidth]{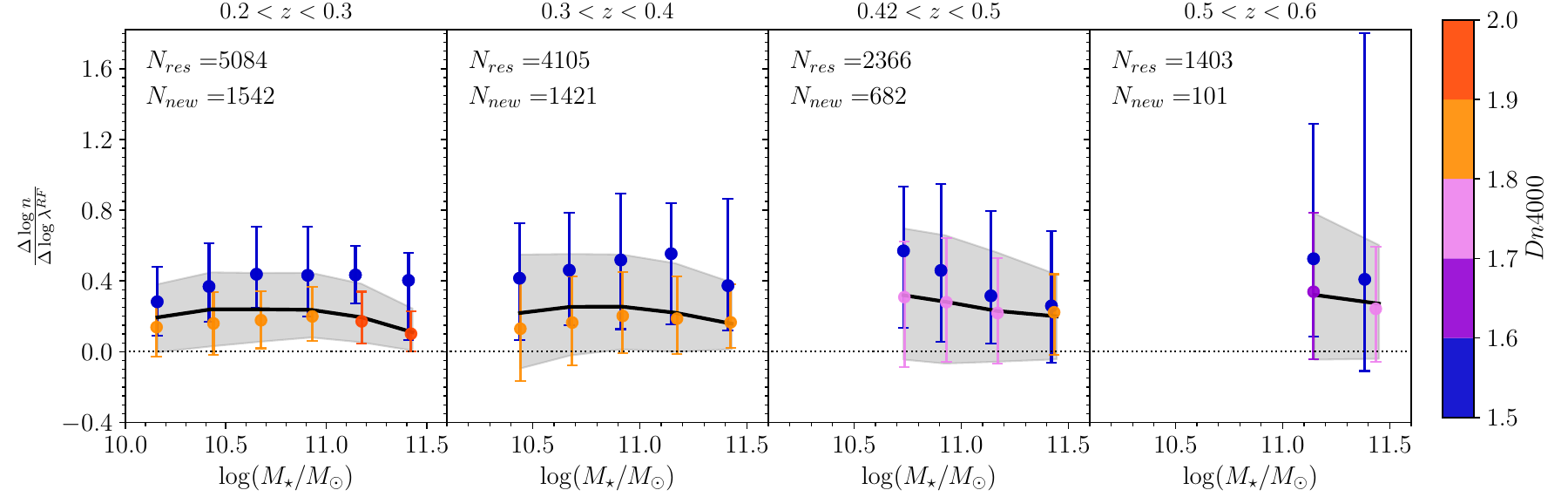}
    \caption{Same as Fig. \ref{fig:gradRdn4000} but for $\frac{\Delta \log n }{\Delta \log \lambda^{RF}}$. }\label{fig:gradndn4000}
\end{figure*}

Figure \ref{fig:gradndn4000} shows the variations in Sérsic index with wavelength as a function of stellar mass for two quiescent populations. At fixed stellar mass, there is a clear decrease in  $\frac{\Delta \log n }{\Delta \log \lambda^{RF}}$ with galaxy global stellar population age. The median rate of increase in S\'ersic index with wavelength for newcomers is $\sim 3, 4, 2$, and 1.5 times larger than for the aging population at $z\sim 0.25, 0.35, 0.45,$ and $ 0.55$, respectively. In all but the highest mass bin at $z<0.4$ (first two panels of Fig.~\ref{fig:gradndn4000}), extremely low $p-$values of the KS tests ($\lesssim10^{-7}$) confirm that this difference in the increase of S\'ersic index with wavelength between two quiescent populations is statistically significant.\footnote{Equivalent to the difference in the wavelength-dependence of size between newcomers and aging population, the observed change in the increase of S\'ersic index with wavelength between the two subpopulations is not statistically significant at $z>0.4$ because the sample sizes and age differences are small.}

At $0.2<z<0.4$, galaxies that only recently ceased star formation exhibit a systematically larger ratio between the central concentrations of their red and UV light profiles with respect to the aging galaxies. Relative to the distribution of their stellar mass, newcomers are less centrally concentrated in UV light than the resident quiescent galaxies that ceased their star formation activity at earlier epochs. 

The changes in morphological parameters with the rest-frame wavelength depend on the stellar population age and thus provide additional constraints on quenching processes. The ratio between the S\'ersic indices in the red and UV rest frames exceeds unity for the newcomers and for the aging population. However, newcomers to the quiescent population show larger ratios than the resident (aging) population in all stellar mass and redshift bins (Fig. \ref{fig:gradndn4000}). At $0.2<z<0.3$, where the difference in age between the old (aging) population and young quiescent galaxies (newcomers) is most significant, the ratio between sizes of galaxies in the red and the UV rest-frame is also systematically larger for the population of newcomers. In this redshift bin, light profiles (i.e., $R_{e,c}$ and $n$) at two rest-frame wavelengths - $3500$ and $7000$~\AA\ - are more similar for the aging population than for the newcomers. 

We explore the effect of $D_n4000$ measurement errors ($\Delta(D_n4000)=0.067$ on average; Sect.~\ref{subsec:hectomap}) on the differences in $R_e$ and $n$ variations with wavelength between newcomers and aging galaxies by repeating the analysis using 1.67 (instead of 1.6) as the upper limit of the $D_n4000$ interval for newcomers. In stellar mass and redshift bins of Figs.~\ref{fig:gradRdn4000} and~\ref{fig:gradndn4000}, the rates of decrease in $R_e$ and increase in $n$  with wavelength for newcomers differ on average by 1.5\% and 4.9\%, respectively. These differences are well within the statistical uncertainties of our $\frac{\Delta \log R_{e,c}}{\Delta \log \lambda^{RF}}$ and $\frac{\Delta \log n}{\Delta \log \lambda^{RF}}$estimates. The variations in structural parameters with wavelength for the aging population are not affected.

The steeper increase in S\'ersic index and decrease in galaxy size with wavelength for the younger quiescent population are consistent with the inside-out quenching scenario \citep[][]{Gonzalez11,Tacchella15}. Spectroscopic samples of quiescent galaxies that include wavelength-dependent measurements of structural parameters and estimates of average stellar population age (e.g., based on D$_n4000$ index) provide improved constraints on the quenching processes and the mechanisms that drive structural evolution in the quiescent phase (Sect.~\ref{subsec:scalingrel}). 

\section{Structural evolution of quiescent galaxies in two rest-frame wavelengths} \label{sect:discussion}

Scaling relations for quiescent galaxies constrain the processes that drive their evolution. We first discuss bias that the observations limited to a single photometric band may introduce in the measured structural parameters of quiescent systems at $0.2<z<0.6$ (Sect. \ref{subsec:error}). We then use rest-frame estimates to explore two scaling relations (size-stellar mass and the S\'ersic index-stellar mass) in two wavelength ranges (UV and red). We trace the corresponding structural evolution in size and central concentration of quiescent galaxies over the $0.2<z<0.6$ redshift interval (Sec. \ref{subsec:scalingrel}). 

\subsection{Sérsic profile parameters in observed and rest-frame wavelengths}\label{subsec:error}

Sizes decrease with wavelength and Sérsic index increase with wavelength for quiescent galaxies with $0.2<z<0.6$ (Sect. \ref{subsec:grad}, Figs. \ref{fig:gradR} and \ref{fig:gradn}). The evolution in galaxy structural parameters is rest-frame wavelength-dependent. The redshift and stellar mass dependence of morphological parameters introduces systematic biases in measurements based on single-band photometric images. These biases influence evolutionary trends inferred from scaling relations. 

\begin{figure}
    \centering
    \includegraphics[width=\linewidth]{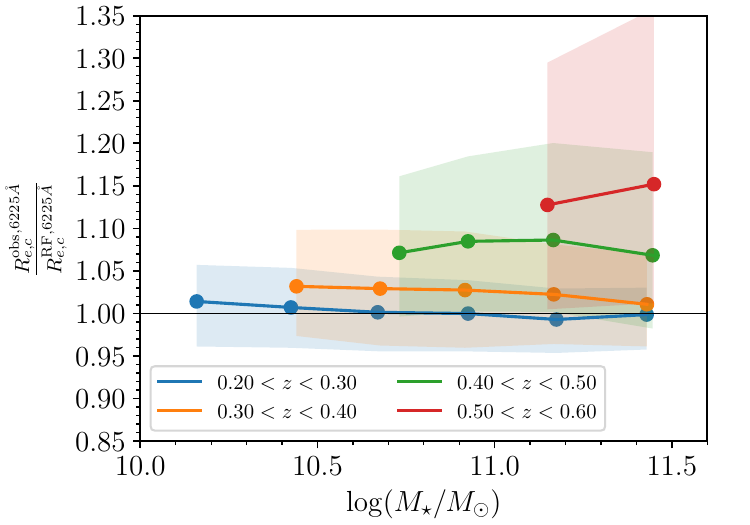}
    \caption{Ratio between sizes measured in the  observed $r$-band  wavelength and in the  rest-frame $r$ band. Points show the median ratios in equally spaced mass bins. Color-coding refers to the four redshift bins in the legend. The shadowed areas indicate the interquartile range of the ratios.  }
    \label{fig:obsrf_r}
\end{figure}

\begin{figure}
    \centering
    \includegraphics[width=\linewidth]{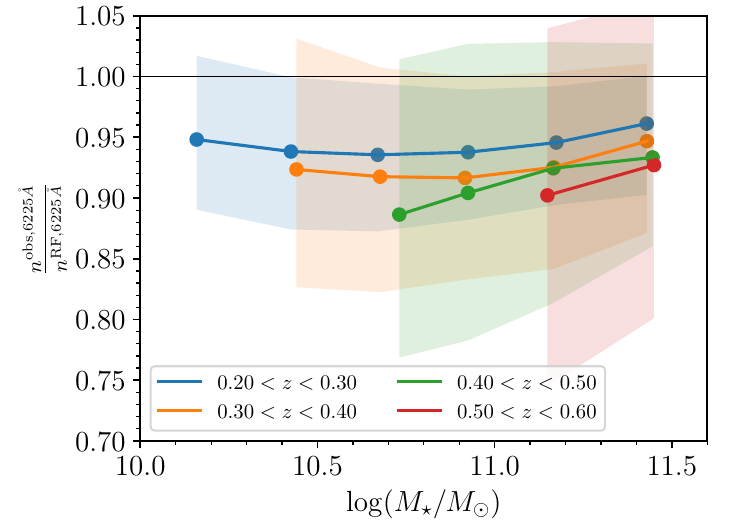}
    \caption{Same as Fig. \ref{fig:obsrf_r}, but the for Sérsic index.}
    \label{fig:obsrf_n}
\end{figure}

In Figs. \ref{fig:obsrf_r} and \ref{fig:obsrf_n}, we quantify the bias in galaxy size and Sérsic index if the changes with wavelength are ignored.
The points in Fig. \ref{fig:obsrf_r} show the median ratio between individual sizes measured in the observed  $r$ band ($R_{e,c}^{{\rm obs},6225\mathring{A}}$) and in the rest-frame $r$ band  ($R_{e,c}^{{\rm RF},6225\mathring{A}}$). The median ratios increase from $\sim 1$ to $\sim 1.15$ as the redshift increases from $z= (0.2-0.3)$ to $z=(0.5-0.6)$.  There is no significant mass dependence.  Sizes are smaller  at longer wavelengths (Fig. \ref{fig:gradR}): $R_{e,c}^{{\rm obs},6225\mathring{A}}$ is systematically larger than $R_{e,c}^{{\rm RF},6225\mathring{A}}$ and the difference increases with increasing redshift. At $z>0.4$ the median bias is significant, $\sim+(10-15)\%$. The interquartile ranges show a large relative scatter of $\sim100\%$  independent of both redshift and stellar mass.

Figure \ref{fig:obsrf_n} shows the ratio between $n^{{\rm obs},6225\mathring{A}}$ and $n^{{\rm RF},6225\mathring{A}}$. The median ratio decreases from $\sim 0.95$ to $\sim 0.90$ as the redshift increases from $z\sim0.25$ to $z\sim0.55$. There is no significant mass dependence. Galaxies are more centrally concentrated at longer wavelength (Fig.~\ref{fig:gradn}). Throughout the redshift range, $n_{e,c}^{{\rm obs},6225\mathring{A}}$ is more than $\sim 5\%$ smaller than $n_{e,c}^{{\rm RF},6225\mathring{A}}$. The median bias  is $\sim-(5-10)\%$. The interquartile ranges in all redshift and stellar mass bins have a constant relative scatter of $\sim 100\%$.

Figures \ref{fig:obsrf_r} and \ref{fig:obsrf_n} show that observations of size or S\'ersic index evolution using a fixed photometric band introduce a bias that slightly enhances any intrinsic growth in size and decreases the change in S\'ersic index. This bias is currently only a small fraction of the error budget associated with the estimates of average change in galaxy structural parameters with redshift (Sec.~\ref{subsec:scalingrel}). Upcoming studies based on large spectroscopic samples will decrease these uncertainties to levels comparable with the bias. Future analyses must account for the single-band photometric biases in structural parameters in order to reliably trace galaxy structural evolution.  

\subsection{Scaling relations for galaxy structural parameters at two different rest-frame wavelengths}\label{subsec:scalingrel}

Next, we examine how galaxy scaling relations depend on the rest-frame wavelength. We discuss the well-known scaling relation between galaxy size and stellar mass (Sect. \ref{subsubsec:rm}) and explore the relation between Sérsic index and mass (Sect. \ref{subsubsec:nm}). To further constrain the processes driving the evolution of quiescent systems, we study the growth in size and the change in S\'ersic index in at rest-frame  UV and red visible wavelengths.  

\subsubsection{Size-stellar mass relation}\label{subsubsec:rm}

The size–mass relation is a fundamental ingredient for exploring evolutionary trends in the quiescent galaxy population. Previous investigations demonstrate that the slope of the relation is essentially redshift independent
at $z<1.5$ \citep[e.g.,][]{Newman12,vdWel14,Faisst17,Mowla19,Barone22}. As the universe ages, the normalization of the relation increases. The increase reflects the size growth of the population \citep[e.g.,][]{Daddi05,Trujillo07,vandokkum10,Damjanov11,Newman12,Huertasc13,vdWel14,Huertasc15,Faisst17,Damjanov19,Mowla19,Kawinwanichakij21,Hamadouche22,Damjanov23,George24,EQ1morpho25}.

The size-stellar mass relation is a function of galaxy stellar population age (traced by $D_n4000$ index). Using quiescent galaxies at $z<0.08$ from SDSS, \citet{Zahid17} show that older objects with higher $D_n4000$ measurements have smaller sizes. A dense spectroscopic survey spanning the redshift interval  $0.2<z<0.5$  (SHELS F2, \citealp{Geller14,Geller16}) confirms the anticorrelation between galaxy size (from HSC-SSP $r$-band imaging) and $D_n4000$ at intermediate redshift \citep{Damjanov18}. \citet{Damjanov23} extends this result to $z\sim 0.6$ using an order of magnitude larger statistical sample of HectoMAP quiescent galaxies with $i$-band imaging. Here we take the advantage of the multiwavelength HSC-SSP coverage of the HectoMAP survey to explore constraints on the size-stellar mass relation for quiescent galaxies in two different rest-frame wavelengths. We examine four redshift bins covering the $0.2 < z < 0.6$ range. 

\begin{figure*}
    \centering
    \includegraphics[width=\linewidth]{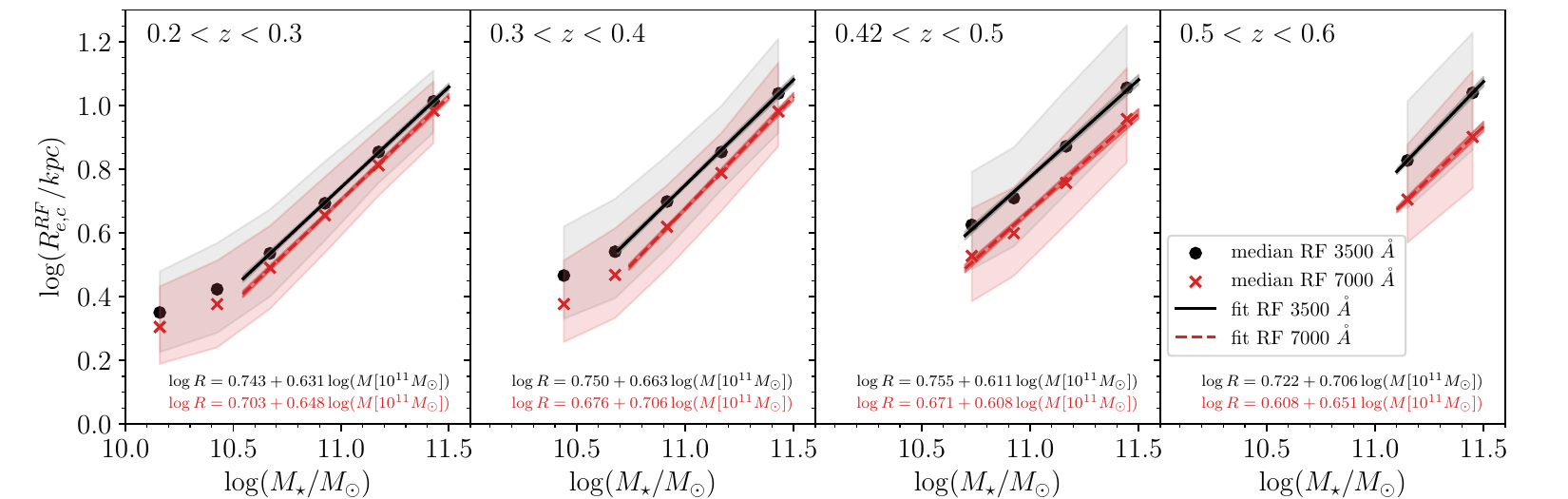}
    \caption{Size-stellar mass relation for quiescent galaxies in two rest frames. The black points and red crosses show the median galaxy size (log scale) for the central values of equally spaced stellar mass bins in the two rest frames $3500$~\AA\ and $7000$~\AA\, respectively. The panels show four redshift bins with redshift increasing from left to right. The black solid and red dashed lines show the linear fit to the medians, and the thickness of dark gray and red bands around them corresponds to the uncertainties. The shaded bands represent the interquartile ranges of galaxy sizes across stellar mass bins.}
    \label{fig:rmrel}
\end{figure*}

Figure \ref{fig:rmrel} shows the size–stellar mass relation for the mass-limited sample of HectoMAP quiescent objects with size estimates at two different rest-frame wavelengths. In each panel, black points and red crosses show the median sizes in 0.25~dex-wide bins of stellar mass at the rest-frame wavelengths of $3500$~\AA\ (hereafter, ``UV'') and $7000$~\AA\ (hereafter, ``red''), respectively. 
We fit the relation $\log R_{e,c}^{RF} = 
R_0^{RF}+ \gamma \log{(M_\star/10^{11}\text{M}_\odot)}$ to the median values. The black solid and red dashed lines show the results in the UV and red rest frames, respectively. 

The size-stellar mass relation for quiescent galaxies exhibits a pivot point at $M_\star \sim (3-5)\times10^{10}$M$_\odot$. The slope of the relation is shallower for masses below the pivot point and steeper for larger masses \citep[e.g.,][]{Shen03,Williams10,Mowla19,Kawinwanichakij21,Nedkova21,Damjanov23,Cutler24,George24,Hamadouche25}. In the two lowest redshift bins  (first two panels from the left), the lower limit of the stellar mass lies below the pivot point mass. For these two redshifts, we fit the median values only for bins with stellar masses exceeding the pivot point mass. We determine the pivot point by using the software package \texttt{pwlf} \citep{pwlf} to fit a continuous piecewise linear function limited to two segments with different slopes. The crossing point of these two segments defines the pivot point mass. In Fig. \ref{fig:rmrel}, the lower extreme of each fitted line identifies the pivot point. At $z>0.4$ the lower limit of the sample completeness is $\sim 1$~dex higher than the completeness limit at $z<0.4$. For $z>0.4$ (two rightmost panels of Fig.~\ref{fig:rmrel}), all of the objects in the samples have stellar masses exceeding the pivot point mass.

The slopes of the size - stellar mass relation for HectoMAP galaxies over the full redshift interval of the survey ($0.2<z<0.6$) are $\gamma_\mathrm{UV}=0.635\pm 0.009$ and $\gamma_\mathrm{red}=0.654\pm 0.009$ in the UV and red rest frame, respectively. Studies based on the Hubble Space Telescope (HST) imaging and spectroscopy \citep{vdWel14,Mowla19} cover $0.2<z<0.5$ with a single redshift bin containing $\sim 100$ galaxies. In both rest-frame regimes, the HectoMAP size-stellar mass relation agrees (within $\lesssim1.5\, \sigma$) with the size-stellar mass relation slope of $0.75\pm 0.06$ from the HST-based study in the $5000$~\AA\ rest-frame \citep{vdWel14}. The value of $\gamma_\mathrm{red}$ is identical to the slope of the size - stellar mass relation for the HectoMAP galaxies at $0.2<z<0.3$ with size measurements based on a single $i$-band HSC imaging \citep{Damjanov23}. 

The zero point of the size-stellar mass relation $R_0^{RF}$, equivalent to the characteristic size of quiescent galaxies at the fiducial stellar mass, is also consistent with previously reported values based on size measurements from either single rest-frame or single observed frame bands. In the redshift range $0.2<z<0.5$, \citet{vdWel14} reports $\log(R_0^{RF, 5000~\AA}/1\, \mathrm{kpc})=0.60\pm0.02$ for a typical $5\times 10^{10}$M$_\odot$ galaxy. At the same stellar mass, the characteristic sizes we measure are $\log(R_0^\mathrm{UV}/1\, \mathrm{kpc})=0.558\pm0.007$ and $\log(R_0^\mathrm{red}/1\, \mathrm{kpc})=0.50\pm0.01$ in the UV and red rest frame, respectively. The characteristic size of HectoMAP galaxies in UV is thus consistent (at $\lesssim2\sigma$ level) with the HST-based study at the rest-frame $5000$~\AA. In  the $0.2 < z < 0.6$ interval, the characteristic size of a $10^{11}\, M_\odot$ HectoMAP galaxy is $\log(R_0^\mathrm{UV}/1\, \mathrm{kpc})=0.743\pm 0.007$ and $\log(R_0^\mathrm{red}/1\, \mathrm{kpc})=0.664\pm 0.006$ in the UV and red rest frame, respectively. The value based on a single ($i$-band) measurement, $\log(R_0/1\, \mathrm{kpc})=0.720\pm 0.004$ \citep{Damjanov23}, lies between the two rest-frame estimates. 

The high density of the HectoMAP survey enables the segregation of quiescent galaxies into narrow bins of redshift and stellar mass. In combination with the multi-band HSC images of the survey area, the HectoMAP survey enables the examination of differences in the redshift evolution of the size-stellar mass relation at two rest-frame wavelengths. In the red rest frame (i.e., at $7000$~\AA, red crosses and dashed lines in Fig. \ref{fig:rmrel}), the zero point of the size-mass relation increases from $\log(R_0^\mathrm{red}/1\, \mathrm{kpc})=0.608\pm 0.007$ to $\log(R_0^\mathrm{red}/1\, \mathrm{kpc})=0.703\pm 0.005$ ($\sim 26\%$) from $z\sim 0.55$ to $z\sim 0.25$. The slope of the size-stellar mass relation in the red rest-frame wavelength shows no significant trend with redshift. 

The increase of the zero point of the size-stellar mass relation with increasing redshift quantifies the size growth of quiescent galaxies. The HectoMAP results in the red rest frame are consistent with previous studies at $z\lesssim1$ that use size measurements based on imaging in a similar rest-frame  \citep{Faisst17,Hamadouche22,Damjanov23}. The negligible evolution in the slope of the relation also agrees with previous work \citep[e.g.,][]{Newman12,vdWel14,Faisst17,Mowla19,Barone22}.

In the UV rest-frame ($3500$~\AA) we detect no significant redshift evolution either in the zero point or the slope of the size-stellar mass relation. We confirm previous findings based on purely photometric samples with size estimates in the same rest-frame wavelength regime\citep{George24}.

\begin{figure}
    \centering
    \includegraphics[width=\linewidth]{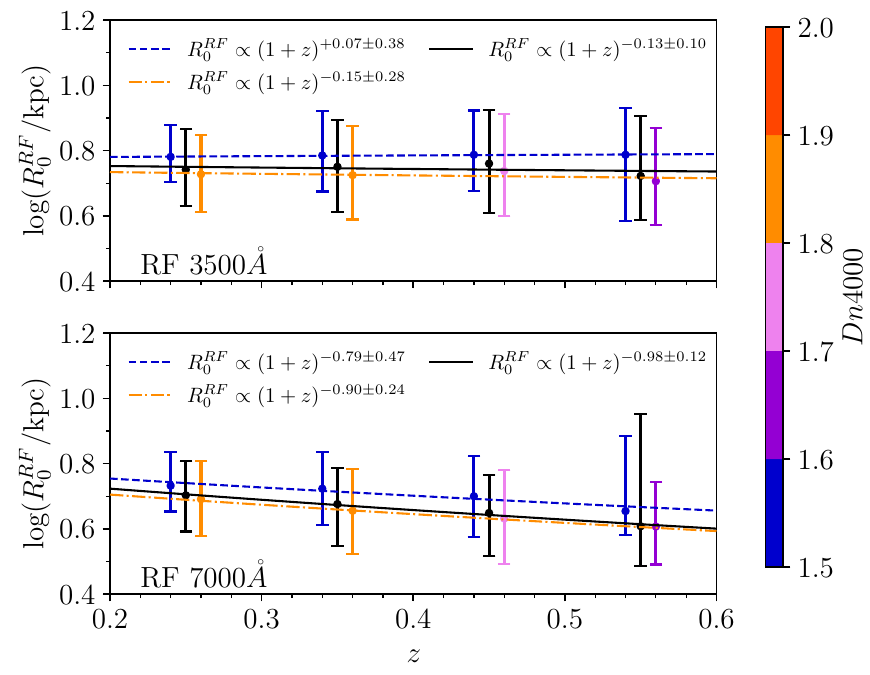}
    \caption{Redshift evolution of the zero point of the size-stellar mass relation (i.e., $\log(R_e/1\, \mathrm{kpc})$ at the reference stellar mass of $10^{11}$~M$_\odot$) as a function of $D_n4000$. The top and bottom panels show the results in the UV and red rest frames, respectively. Black points show the median size of the total sample of galaxies, and the sets of colder and warmer points show the zero points for the subpopulations of newcomers and aging galaxies, respectively. The colors of symbols for the two subpopulations correspond to their median $D_n 4000$. The error bars denote the 25th-75th percentile range of the distribution. The dashed blue and dash-dotted orange lines are power law fits to the evolution of the zero point of the subpopulations of newcomers and aging galaxies, respectively.
    }
    \label{fig:rDNevo}
\end{figure}
Figure \ref{fig:rDNevo} illustrates the size evolution of a typical $10^{11}\, M_\odot$ HectoMAP galaxies with different average stellar population ages at two rest-frame wavelengths. To derive the characteristic sizes in the first three redshift bins, we interpolate between the median sizes of the two mass bins with mass nearest to $10^{11}$M$_\odot$. We compute the characteristic size as the interpolated value at this fiducial mass. The interpolated values are equivalent to the values of zero points $R_0$ from the best-fit linear relations in Fig. \ref{fig:rmrel}. In the highest redshift bin, the characteristic size is the extrapolation of the size - stellar mass relation defined by the two points at $M_\star>10^{11}\, M_\odot$. 

The black circles with error bars in Fig. \ref{fig:rDNevo} show the redshift evolution in the characteristic size for the full quiescent sample and its scatter in the UV (upper panel) and red (bottom panel). The black lines show the best-fit relation of the form $R_0^{RF}(z)= a(1+z)^b$. The slope of the relation in the red rest frame is $-0.98\pm 0.11$, much steeper than the slope in the UV, $-0.13\pm 0.10$. 
As redshift decreases from $z\sim 0.55$ to $z\sim 0.25$, the ratio between the characteristic sizes in UV and red decreases from $\sim 1.3$ to $\sim 1.1$.

Figure~\ref{fig:rDNevo} also explores the redshift evolution in galaxy sizes in two rest-frame wavelengths for two quiescent subpopulations based on $D_n4000$. We select newcomers (blue colored points) and the aging population (purple/red shaded color points) following the procedure described in Sect. \ref{subsec:grad_dn4000}. 

The change in characteristic size with redshift is similar for the newcomers and for the aging population. In the UV rest-frame, the sizes of two subpopulations are essentially constant over the range $0.2<z<0.6$. For the red rest frame wavelength, the sizes of newcomers and aging galaxies increase by $\sim 26-30\%$ from the high-redshift to the low-redshift limit. At fixed mass and redshift, newcomers are larger than aging galaxies in both rest frames. Characteristic sizes for the total sample always lie between the sizes of these two subpopulations. The uncertainties in $D_n4000$ measurements (Sect.~\ref{subsec:grad_dn4000}) have no impact on these results. The offset in size between young quiescent galaxies (newcomers) and their older counterparts (aging population) and the size growth in the rest-frame red light agrees with previous studies based on single-band (visible red or near-infrared) imaging \citep[e.g.,][]{Zahid17,Hamadouche22,Damjanov23}.

Minor mergers are widely regarded as the main driver of quiescent galaxy size evolution at $z\lesssim1$ \citep{Naab09,Nipoti09,Nipoti12,Newman12,Faisst17,Hamadouche22,Damjanov24}. 
In this scenario, the absence of size growth in the rest-frame UV for massive ($M_\star\sim10^{11}\, M_\odot$) aging galaxies (purple and red colored points in the top panel of Fig. \ref{fig:rDNevo}) may indicate that the star forming material and/or metal-poor stars are mainly accreted at $z\gtrsim0.5$ where minor mergers with blue satellites are more frequent \citep[e.g.,][]{Lotz11,Suess23}. At $z\lesssim0.55$, accreted material is consumed in residual star formation episodes in the galaxy outskirts. The metal-poor stars accreted at higher redshift contribute to the galaxy halo light. The constant size of massive newcomers in UV light mirrors the lack of substantial size growth in UV for their massive star-forming counterparts \citep{George24}. 

The size increase with decreasing redshift in the red rest frame (bottom panel of Fig. \ref{fig:rDNevo}) implies that at lower redshift the distribution of the older stellar population (i.e., the bulk of stellar mass) extends to larger radii. This size growth in red light for the aging population suggests an increase in the fraction of low-mass red (i.e., old and/or metal-rich) satellites merging with massive quiescent galaxies at later epochs.  Observing the color evolution of low-mass satellites at intermediate redshift is challenging due to their surface brightness \citep{Guo11,Nierenberg13}. With the promised increase in imaging sensitivity and survey area,  developing and ongoing missions (e.g., \textit{Euclid}, \citealt{Mellier25}; \textit{Vera Rubin} Observatory's LSST, \citealt{Ivezic2019}) may soon provide large enough samples of low surface brightness systems to test this hypothesis. 

Residual star formation has a similar effect on the size growth of both newcomers and aging quiescent galaxies in the red rest frame. This process adds stellar mass to the galaxy outskirts, increasing  galaxy size in the red rest frame. Regions dominated by blue light are at galactocentric distances larger than the regions encompassed by the Hectospec fibers at intermediate redshift \citep[$0.5-2~R_e$ in observed $i$ band,][]{Damjanov24} and thus they do not affect the global average stellar population age derived in the HectoMAP survey. Spatially resolved spectroscopic observations are critical for measuring stellar population age (and metallicity) in galaxy outskirts. However, these low surface brightness regions are beyond the reach of currently available integral field unit (IFU) observations in the rest-frame UV/visible light even for $z\sim0$ galaxies \citep[the limits are $<1.5-2\, R_e$,][]{Lacerna20,Parikh21,Shengdong23}.  

Based on the combination of observed size growth in the UV and in the red, a low level of star formation in galaxy outskirts may be sufficient to account for the structural evolution of $M_\star\sim10^{11}\, M_\odot$ newcomers. The residual star formation should be at a lower level in the outskirts of aging quiescent galaxies of the same stellar mass. Thus minor mergers with red satellites are necessary to account for the observed size evolution of the aging quiescent subpopulation.

\subsubsection{Sérsic index-stellar mass relation}\label{subsubsec:nm}

The scaling relation between S\'ersic index and stellar mass ($n-M_\star$ relation) probes the effects that the processes driving the mass assembly of quiescent systems have on the central concentration of their light profiles. 
Small galaxy samples and significant uncertainties in individual measurements of $n$ limit the number of studies that address the $n-M_\star$ relation \citep[e.g.,][]{vandokkum10,Huertasc13}. These studies cover a broad redshift interval ($0<z<2$) and find heterogeneous trends with redshift. The results in the literature include both a decrease and a slight increase of $n$ with increasing redshift. Most recent results based on large photometric samples from JWST show that the Sérsic index of quiescent galaxies with $\log (M_\ast/M_\odot)>8$ slightly increases from $n\sim 4.5$ at $z\sim 2$ to $n\sim 5.4$ at $z\sim 0.5$ \citep{Hamadouche25}.

With the dense HectoMAP sample we explore the evolution in $n-M_\star$ relation as a function of the rest-frame wavelength used for the estimates of structural parameters. We also explore the dependence of this relation on the global stellar population age that separates newcomers from the aging subpopulation.

\begin{figure*}
    \centering
    \includegraphics[width=\linewidth]{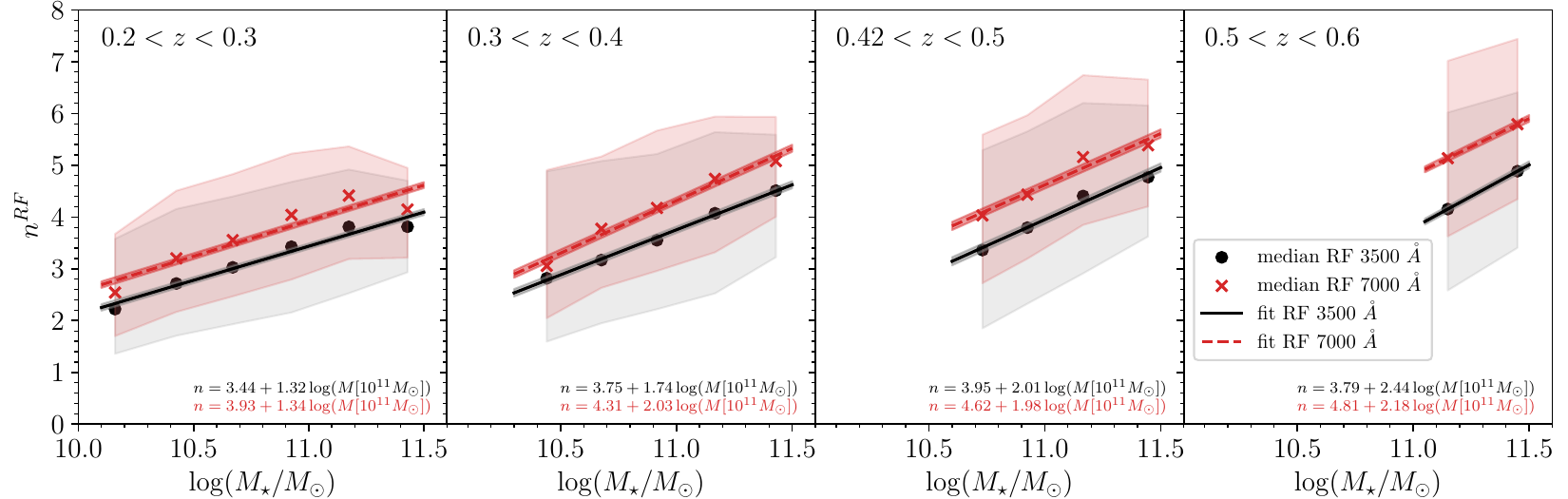}
    \caption{Sérsic index-stellar mass relation for quiescent galaxies in two rest frames. Analogously to Fig. \ref{fig:rmrel}, the black points and red crosses show the median galaxy concentration for the central values of equally spaced stellar mass bins in the two rest frames, UV and red, respectively, in four redshift bins with redshift increasing from left to right. The black solid and red dashed lines show the linear fit to the medians, and the thickness of dark gray and red bands around them corresponds to the uncertainties. The shaded bands represent the interquartile ranges of galaxy Sérsic indices across stellar mass bins. }
    \label{fig:nmrel}
\end{figure*}

Figure \ref{fig:nmrel} shows the $n-M_\star$ relation for  HectoMAP quiescent galaxies in two different rest frames. In each panel, black points and red crosses show the median sizes in 0.25~dex-wide bins of stellar mass in the UV and red rest frame, respectively. 
We fit the relation $n^{RF} = n_0^{RF} + \gamma \log{(M_\star/10^{11}\text{M}_\odot)}$ to the median values. The black solid and red dashed lines show the results in UV and red rest frame, respectively. 

The slope of the $n-M_\star$ relation increases with redshift in both rest frames. From $z\sim 0.25$ to $z\sim 0.55$, the slope increases by $\sim 63\%$ and $\sim 85\%$ in the UV and red rest frames, respectively. The Sérsic indices in the UV are systematically smaller than those in red; the offset increases with increasing redshift.
The zero point of the $n-M_\star$ relation $n_0^{RF}$, the characteristic Sérsic index of a $10^{11}$M$_\odot$ quiescent galaxy, shows stronger evolution in the red than in the UV rest-frame. In the UV rest-frame, the zero point increases by $\sim 10\%$ (around $n_0^{\rm UV}\sim3$) as the redshift increases from $z\sim 0.25$ to $z\sim 0.55$. In the red rest frame the zero point  increases from $n_0^{\rm red}=3.93\pm 0.05$ to $n_0^{\rm red}=4.81\pm 0.06$ ($\sim 22\%$) over the same redshift range.

\begin{figure}
    \centering
    \includegraphics[width=\linewidth]{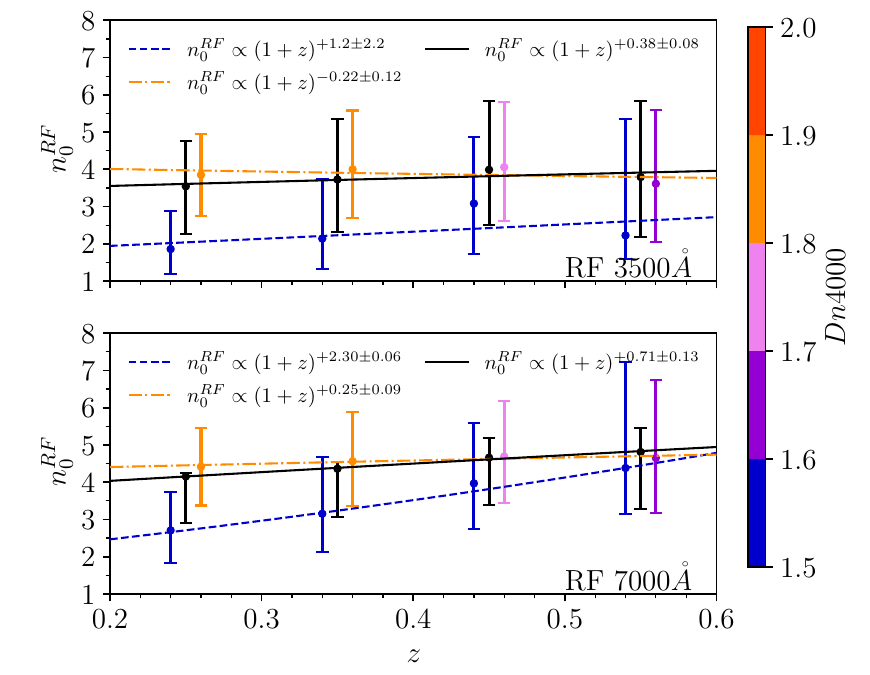}
    \caption{  Redshift evolution of the zero point of the Sérsic index-stellar mass relation (i.e., $n$ at the reference stellar mass of $10^{11}$~M$_\odot$) as a function of $D_n4000$. Analogously to Fig. \ref{fig:rDNevo}, the top and bottom panels show the results in the UV and red rest frames, respectively. Black points show the median Sérsic index of the total sample of galaxies, and the sets of colder and warmer points show the zero points for the subpopulations of newcomers and aging galaxies, respectively. The error bars denote the interquartile range of the distribution. The dashed blue and dash-dotted orange lines are power law fits to the evolution of the zero point of the subpopulations of newcomers and aging galaxies, respectively. }
    \label{fig:nDNevo}
\end{figure}
We use interpolated values for the S\'ersic index at $M_\star=10^{11}\, M_\odot$, based on the medians of measured $n$ values in the adjacent mass bins, as the characteristic S\'ersic indices $n_0^\mathrm{UV}$ and $n_0^\mathrm{red}$ in two rest frames and show their evolution for the complete quiescent sample in Fig. \ref{fig:nDNevo}. The black circles with error bars display the redshift evolution of the characteristic S\'ersic index and its scatter in the UV (upper panel) and red (lower panel) rest-frame. The black lines show the fit of the characteristic indices to the relation $n_0^{RF}(z)= a(1+z)^b$. The exponent of the relation in the red rest frame is $b=+0.71\pm 0.13$, significantly steeper than the exponent in the UV ($b=+0.38\pm 0.08$). 

We also explore the redshift evolution of the characteristic S\'ersic index at two rest-frame wavelengths as a function of the average $D_n4000$  for the populations of newcomers and aging galaxies (Sect. \ref{subsec:grad_dn4000}, color-coded points in Fig. \ref{fig:nDNevo}). In the UV rest-frame $M_\star\sim10^{11}\, M_\odot$ newcomers are less centrally concentrated (i.e., more disk-like with $n_0^{\rm UV}\sim2$) than the spheroidal ($n_0^{\rm UV}\sim4$) aging quiescent galaxies. The two subpopulations show essentially constant $n_0^{\rm UV}(z)$. The evolution in the S\'ersic index measured in rest-frame UV light for newcomers and aging galaxies is consistent with the redshift trend for the total sample.

In the red rest frame, the characteristic index $n_0^\mathrm{red}$ for both newcomers and aging galaxies increases with increasing redshift, albeit at different rates. For the aging subpopulation, the central light concentration increases negligibly with redshift. In contrast, the drop in the central concentration of newcomers is significant. At $z\sim 0.25$, newcomers have $\sim 2$ times smaller $n$ than aging galaxies; at $z\sim 0.55$, the two subpopulations exhibit similar $n$ values. Equivalently to the characteristic size $R_0^{RF}$ (Sect. \ref{subsubsec:rm}), the evolutionary trends in $n_0^{RF}$ at two rest-frame wavelengths for newcomers and aging population are not sensitive to the uncertainties in $D_n4000$ measurements.

A relatively constant value of $n$ for $M_\star\sim10^{11}\, M_\odot$ quiescent galaxies in the rest-frame UV complements the lack of evolution in size $R_0^\mathrm{UV}$ of quiescent galaxy light profiles (Fig.~\ref{fig:rDNevo}). The offset in the rest-frame UV  values of $n$ between newcomers and the aging population illustrates the difference in the concentration of the blue light profiles. For newcomers, these light profiles are disk-like. In contrast, the blue light regions of aging quiescent galaxies are spheroidal halos. 

The change with redshift in the characteristic $n$ for the aging population is very mild in the rest-frame red. The trend is consistent with no evolution within $2.75\, \sigma$. Furthermore, the values remain very close to $n=4$, a classical de Vaucoulers \citep{deVacouleurs48} profile for elliptical galaxies dominated by an old stellar population. 

For the subpopulation of newcomers characteristic S\'ersic index increases substantially with redshift in the red rest frame, from  $n\sim2.5$ at $z\sim0.25$ to $n\sim4.5$ at $z\sim0.55$. This evolutionary trend implies that the distribution of the bulk of stellar mass in star-forming galaxies at $0.6<z<0.7$ (progenitors of newcomers at $z\sim0.55$) is much more centrally concentrated than at $\Delta z\sim0.2$ lower redshift. 

The significant apparent change in $n_0^\mathrm{red}$ for newcomers may arise from observational effects. The Hectospec fiber encloses a $\sim 50\%$ smaller central region at $z\sim 0.2$ than at $z\sim 0.6$ \citep{Damjanov24}. IFU-based studies of $z<0.15$ galaxies show that the radial decrease in stellar population age is most prominent for galaxies in between the star-forming sequence and quiescent galaxies \citep{Shengdong23}. As the redshift decreases, the combination of fixed angular fiber size and the strong radial age gradient can bias the apparent properties of newcomers towards lower D$_n4000$. This selection effect results in the increasing prominence of disk-like systems ($n\lesssim2$) in the lower-redshift samples and explains the unusual apparent evolutionary trend we observe for newcomers. This effect should have no impact on the aging quiescent population because the radial age profiles in galaxies with globally old stellar population are flat \citep{Shengdong23}.

The scaling relation between S\'ersic index $n$ and galaxy stellar mass in the UV and red rest frames shows that all quiescent systems have blue light profiles that are less centrally concentrated than their light profiles at redder wavelengths. When separated into subpopulations of recently quenched systems (newcomers) and the oldest resident (aging) population based on their $D_n4000$ index, UV rest-frame profiles of young quiescent galaxies are, on average, disk-like, and the profiles of aging population in the same wavelength regime are spheroidal. In contrast with the size evolution in the red rest frame (Sect.~\ref{subsubsec:rm}), the evolution in S\'ersic index is very mild and, within the uncertainties, indicates little or no variation in the light profiles between $z\sim0.55$ and $z\sim0.25$ in both rest frames. Together with larger $n_0^{RF}$ for the aging subpopulation in both rest frames, the relatively constant central concentration of light profiles at $0.2<z<0.6$ points at the residual (and/or preceding) star formation (newcomers) and minor mergers (aging subpopulation) as the main drivers of the observed structural evolution of quiescent systems at intermediate redshifts. 

\section{Conclusion}\label{sec:conclusion}

Large (complete) mass limited samples of quiescent galaxies provide an increasingly detailed picture of their evolution. The HectoMAP survey \citep{Geller11Russel, Sohn21HDR1, Sohn23, Damjanov23}, combines MMT/Hectospec spectroscopy and Subaru/HSC five-band imaging to yield a mass-limited sample of 26,912 quiescent objects with structural parameter measurements spanning the redshift range $0.2 <z <0.6$. The spectra provide measurements of galaxy redshift and spectral index D$_n4000$, the indicator of average stellar population age. HSC-SSP DR3 five-band imaging enables measurements of morphological parameters at the rest-frame wavelengths $(3500-7000)$~\AA\ for 80\% of the total area covered by HectoMAP. We used a combination of photometry and spectroscopy to derive stellar masses. The HectoMAP quiescent sample is 6.6 times the size of the largest comparable sample previously available at these redshifts with single-band imaging \citep[SHELS,][]{Geller14,Geller16}. 

We fit a single S\'ersic model to the five-band images of the quiescent galaxy sample (Sect. \ref{subsec:ress}) and used the best-fit linear relation between structural parameters and the wavelength of corresponding images to measure half-light radius and S\'ersic index in two rest frames: $3500$~\AA\ (UV) and $7000$~\AA\ (red). For each galaxy, we measured the ratio between the logarithmic difference in the circularized half-light radius or S\'ersic index and the logarithmic difference in the two rest-frame wavelengths of the two measurements (Sect. \ref{subsec:grad}).

The dense HectoMAP spectro-photometric sample enables an exploration of the evolution of the change in structural parameters with the rest-frame wavelength over four equally spaced redshift bins covering the range  $0.2<z<0.6$  (Sect. \ref{subsec:grad}). Galaxy size decreases with rest-frame wavelength, independent of mass. The strength of this anti-correlation grows weakly with redshift. In contrast, the Sérsic index increases with rest-frame wavelength. The amplitude of this increase is essentially independent of mass and redshift. The distribution of the bulk of stellar mass in quiescent systems (traced by redder rest-frame wavelength) is more centrally concentrated and less extended than the distribution of younger and/or metal-poor stellar populations dominating galaxy light profiles in bluer rest-frame wavelengths. At intermediate redshift, the HectoMAP survey provides the measurements of structural parameters that cover a broad rest-frame wavelength range, previously available for statistically large spectroscopic samples only at $z<0.2$ \citep[e.g.,][]{Kelvin12,Kennedy16}. 

Using $D_n4000$ as a proxy for galaxy average stellar population age we investigate changes in the wavelength-dependence of morphological parameters with galaxy age. We consider two subpopulations of the quiescent sample: newcomers that only recently joined the quiescent populations, and the aging subpopulation that ceased their star formation at $z>0.6$ (Sect.~\ref{subsec:grad_dn4000}). 

In the lowest redshift bin where the samples of these two quiescent subpopulations are largest,  the size decreases with rest-frame wavelength significantly faster for newcomers than for the aging subsample. Furthermore, the increase in S\'ersic index with rest-frame wavelength is systematically larger for newcomers than for aging galaxies in all stellar mass and redshift bins. Newcomers to the quiescent population exhibit a stronger dependence on the rest-frame wavelength than the resident (aging) population in both size and Sérsic index, further supporting the prevalence of inside-out quenching at $z<0.6$.

We study the size-stellar mass and S\'ersic index-stellar mass relation for quiescent population in the UV and red rest frames (Sects. \ref{subsubsec:rm} and \ref{subsubsec:nm}). In the red rest frame, the characteristic size of $10^{11}$M$_\odot$ quiescent galaxies increases with decreasing redshift as $R_0^\mathrm{red}\propto (1+z)^{-0.98\pm 0.11}$, or $\sim 30\%$ from $z\sim0.55$ to $z\sim0.25$. This result broadly agrees with previous investigations based mostly on size measurement in a single (red) observed or rest-frame photometric band  \citep{vdWel14,Mowla19,Damjanov23}. On the other hand, in the rest-frame UV, the change in the characteristic size is much slower and consistent with no evolution. As the redshift increases, the typical S\'ersic index of a $10^{11}$M$_\odot$ quiescent galaxy increases faster at longer wavelengths: it increases (modestly) by $\sim 7\%$ and $\sim 14\%$ in the UV and red rest frames, respectively, from $z\sim 0.2$ to $z\sim 0.6$.

 The newcomers and aging subpopulation both show nearly constant (rest-frame) UV-based characteristic size and S\'ersic index over the redshift range $0.2<z<0.6$, with the newcomers displaying an offset towards slightly ($\sim20\%$) larger size and two times lower S\'ersic index. In the red rest frame, the characteristic sizes of both subpopulations increase at the same rate (within the uncertainties) as the size of a $M_\star=10^{11}\, M_\odot$ galaxy representing the complete quiescent sample. Conversely, a decrease in the S\'ersic index with cosmic time is negligible for the resident population and significant for the newcomers ($\sim42\%$ from $z\sim0.6$ to $z\sim0.2$). This difference arises because a fixed spectrograph fiber size, combined with strong color gradients and size evolution in newcomers, preferentially selects younger newcomers at lower redshifts (Sect.~\ref{subsubsec:nm}).

Minor mergers are widely regarded as the main channel for growth of individual quiescent galaxies \citep[e.g.,][]{Naab09,Nipoti09,Newman12,Faisst17,Hamadouche22,Damjanov24,Hamadouche25}. In this context, the absence of change in the structural properties of the aging population in the rest-frame UV suggests that the star-forming-and-low-metallicity material is deposited into the extended regions of galaxy light profiles through minor mergers. As the merger frequency decreases with cosmic time, this material maintains a low level of residual star formation and settles into galaxy halos. On the other hand, size growth in the red rest frame  is driven by a combination of added stellar mass (through residual star formation) and a higher frequency of red satellites merging with massive quiescent galaxies at later epochs. 

The HectoMAP quiescent sample demonstrates the power of a large, complete mass-limited quiescent sample for exploring evolutionary trends in the extent and shape of light distributions, over a broad baseline of rest-frame wavelengths, for quiescent subpopulations segregated by average stellar population age. Ongoing and planned deep redshift surveys with unprecedentedly large area coverage \citep[e.g., DESI Luminous Red Galaxies, PFS Galaxy Evolution Survey;][]{Zhou23,Greene22} will allow for  stellar population age-based selection of galaxy subpopulations in volume-limited samples up to $z>1$. In parallel, space-based photometric surveys will provide robust multiwavelength measurements of the half-light radii and S\'ersic indices for large samples of quiescent galaxies over a a broad redshift range \citep[e.g., \textit{Euclid};][]{EQ1MER25, EQ1morpho25}, enabling statistical probes of the changes in structural properties with the rest-frame wavelength at higher redshifts. 

The HectoMAP mass-complete spectro-photometric quiescent sample at intermediate redshift is a platform for the first statistical determination of galaxy structural parameters as a function of the rest-frame wavelength at $0.2 < z < 0.6$. Spanning almost half of cosmic history, our analysis of this sample yields critical insights into the dominant processes driving galaxy quenching and structural evolution during quiescence. We have established a foundational template for future surveys to extend these probes to even earlier epochs, thereby laying the groundwork for a more detailed, coherent picture of galaxy mass assembly.

\section*{Data availability}

Table \ref{table:measures} is only available in electronic form at the CDS via anonymous ftp to \texttt{cdsarc.u-strasbg.fr (130.79.128.5)} or via \texttt{http://cdsweb.u-strasbg.fr/cgi-bin/qcat?J/A+A/}.

\begin{acknowledgements}
We thank the anonymous referee for providing insightful comments and suggestions.
M.P. and I.D. acknowledge the support of the Canada Research Chair Program and the Natural Sciences and Engineering Research Council of Canada (NSERC, funding reference number RGPIN-2018-05425). 
This research has made use of NASA’s Astrophysics Data System Bibliographic Services.

The Hyper Suprime-Cam (HSC) collaboration includes the astronomical communities of Japan and Taiwan as well as Princeton University. The HSC instrumentation and software were developed by the National Astronomical Observatory of Japan (NAOJ), the Kavli Institute for the Physics and Mathematics of the Universe (Kavli IPMU), the University of Tokyo, the High Energy Accelerator Research Organization (KEK), the Academia Sinica Institute for Astronomy and Astrophysics in Taiwan (ASIAA), and Princeton University. Funding was contributed by the FIRST program from the Japanese Cabinet Office, the Ministry of Education, Culture, Sports, Science and Technology (MEXT), the Japan Society for the Promotion of Science (JSPS), Japan Science and Technology Agency (JST), the Toray Science Foundation, NAOJ, Kavli IPMU, KEK, ASIAA, and Princeton University. This paper makes use of software developed for the Large Synoptic Survey Telescope (LSST). We thank the LSST Project for making their code available as free software at http://dm.lsst.org. This paper is based [in part] on data collected at the Subaru Telescope and retrieved from the HSC data archive system, which is operated by Subaru Telescope and Astronomy Data Center (ADC) at National Astronomical Observatory of Japan. Data analysis was in part carried out with the cooperation of Center for Computational Astrophysics (CfCA), National Astronomical Observatory of Japan.
\end{acknowledgements}

\bibliographystyle{aa}
\bibliography{biblio}

\begin{appendix}

\onecolumn
\section{Comparison with previous size estimates}\label{app:comparison}

Analysis of \sepp performance guarantees the reliability and the essential absence of bias of our measurements \citep[see Sect. \ref{subsec:sepp}, and][]{Bretonniere23}. Here, we show a sanity check of our results by comparing our findings with independent measurements of morphological parameters derived for the same sample by \citet{Damjanov23}, that use \se to fit a single-Sérsic model to the HSC/SSP $i-$band photometry of the HectoMAP sample. 

\begin{figure*}[h]
    \centering
    \includegraphics[width=\textwidth]{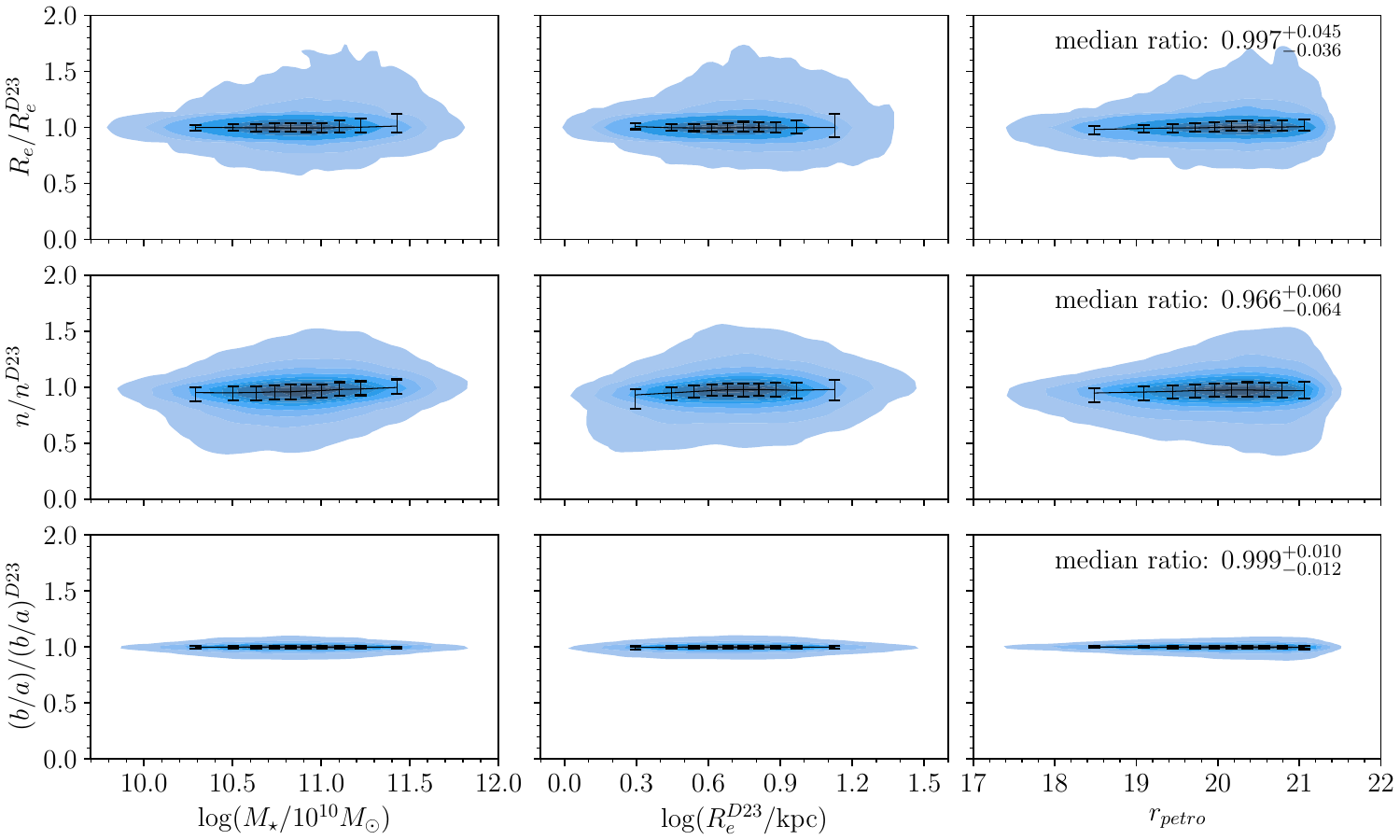}
    \caption{Comparison with previous measurements for HectoMAP galaxies in $i-$band. The upper row shows the ratio between our $R_e$ \sepp measurements and the corresponding size $R_e^{D23}$ obtained by \citep{Damjanov23} using the \se software. The middle and lower rows show the same, but for $n$ and $b/a$, respectively. From left to right, the panels show the dependence on $M_\star$, $R_e^{D23}$, and $r_{petro}$. Blue regions show the density distribution obtained with a Gaussian kernel estimator. The darkest (smallest) region encloses the 10\% of the sample, while the lightest (broadest) region includes the 90\% of the sample; successive levels are separated by equal increases of $+10\%$ of the total sample. The black lines connect the medians of the ratios in ten equally sampled bins of the corresponding $x$-axix. Error bars show the interquartile range of the corresponding bin.   }
    \label{fig:comp_d19}
\end{figure*}

Figure \ref{fig:comp_d19} shows the results for the HectoMAP mass-complete sample of quiescent galaxies with $R_e$ larger than half the median $i$-band seeing ($0.55$~arcsec). From left to right, the upper row shows the ratio $R_e/R_e^{D23}$ between our sizes $R_e$ and the corresponding sizes $R_e^{D23}$ by \citet{Damjanov23} as a function of $M_\star$, $R_e^{D23}$, and $r_{petro}$, respectively. The middle and lower rows show the same for Sérsic index, $n/n^{D23}$, and axis ratio, $(b/a)/(b/a)^{D23}$, respectively. In each panel, black line and error bars show median and the interquartile ranges in ten equally sampled bins of the corresponding $x$-axis.

Figure \ref{fig:comp_d19} proves the absence of bias between our \sepp measurements and \citet{Damjanov23} \se fits. 
Size, Sérsic index, and axis ratio from the two software are consistent with each other and do not show any trends in function of $M_\star$, $R_e^{D23}$, and $r_{petro}$. In every panel and every bin, the interquartile ranges are fully consistent with equality between our measurements and the results of \citet{Damjanov23}. On average, $R_e/R_e^{D23}=0.997$ with a small dispersion $\sim 4\%$, $n/n^{D23}=0.966$ with a $\sim 6\%$ dispersion, and $(b/a)/(b/a)^{D23}=0.999$ with a small $\sim 1\%$ dispersion.

\end{appendix}

\end{document}